\documentclass{aa}  

\usepackage{graphicx}
\usepackage{txfonts}
\usepackage{CJK}
\usepackage{enumitem}
\usepackage{amsmath}
\usepackage{amssymb}
\usepackage{natbib}
\usepackage{hyperref}
\hypersetup{
    colorlinks=true,
    linkcolor=blue,
    filecolor=blue,      
    urlcolor=blue,
    citecolor=blue
}

\newcommand   {\EBV}     {E(B\,{-}\,V)}
\newcommand   {\KI}      {\rm K\,\uppercase\expandafter{\romannumeral1}}
\newcommand   {\NI}      {\rm N\,\uppercase\expandafter{\romannumeral1}}
\newcommand   {\FeI}     {\rm Fe\,\uppercase\expandafter{\romannumeral1}}
\newcommand   {\TiI}     {\rm Ti\,\uppercase\expandafter{\romannumeral1}}
\newcommand   {\CaI}     {\rm Ca\,\uppercase\expandafter{\romannumeral1}}
\newcommand   {\EJKs}    {E(J\,{-}\,K_{\rm S})}
\newcommand   {\JKs}     {J-K_{\rm S}}

\newcommand   {\Av}      {A_{\rm V}}
\newcommand   {\Rv}      {R_{\rm V}}

\newcommand   {\Teff}    {T_{\rm eff}}
\newcommand   {\logg}    {{\rm log}\,g}
\newcommand   {\feh}     {\rm [Fe/H]}

\newcommand   {\kms}     {\rm km\,s^{-1}}

\newcommand   {\um}      {\rm \mu m}

\newcommand   {\Vlsr}    {V_{\rm LSR}}
\newcommand   {\Vhc}     {V_{\rm HC}}
\newcommand   {\Vgc}     {V_{\rm GC}}
\newcommand   {\Usun}    {U_{\odot}}
\newcommand   {\Rsun}    {R_{\odot}}
\newcommand   {\Rgc}     {R_{\rm GC}}

\newcommand   {\dlos}    {d_{\rm los}}

\defcitealias{hz21}{Paper\,$\rm \uppercase\expandafter{\romannumeral1}$}
\defcitealias{SFD}{SFD}
\defcitealias{Surot2020}{S20}
\defcitealias{Alvaro17}{R17}
\usepackage[normalem]{ulem}

\makeatletter
\renewcommand*\aa@pageof{, page \thepage{} of \pageref*{LastPage}}
\makeatother

\begin{document}

\begin{CJK*}{UTF8}{gbsn}

\title{The diffuse interstellar band around 8620 {\AA}}
   
\subtitle{II. Kinematics and distance of the DIB carrier}

\author{H. Zhao (赵赫)\inst{1} 
        \and
        M. Schultheis\inst{1}
        \and
        A. Rojas-Arriagada\inst{2,3}
        \and 
        A. Recio-Blanco\inst{1}
        \and 
        P. de Laverny\inst{1}
        \and
        G. Kordopatis\inst{1}
        \and
        F. Surot \inst{4,5}
        }

\institute{University C\^ote d'Azur, Observatory of the C\^ote d'Azur, CNRS, Lagrange
          Laboratory, Observatory Bd, CS 34229, \\
          06304 Nice cedex 4, France \\
          \email{he.zhao@oca.eu, mathias.schultheis@oca.eu}
          \and
          Instituto de Astrof\'isica, Facultad de F\'isica, Pontificia, Universidad Cat\'olica de Chile, Av. Vicu\~na Mackenna 4860, Santiago,  Chile
          \and
          Millennium Institute of Astrophysics, Av. Vicu\~na Mackenna 4860, 782-0436 Macul, Santiago, Chile
          \and
          Departamento de Astrof\'isica, Universidad de La Laguna, E\--38205, La Laguna Tenerife, Spain
          \and
          Instituto Astrof\'isica de Canarias, V\'ia L\'actea S/N, E-38200 La Laguna, Tenerife, Spain
      }

\date{Received Apr 19, 2021; accepted Aug 27, 2021}


 
\abstract
{Diffuse interstellar bands (DIBs) are important interstellar absorption features of which the origin is still debated. With the 
large data sets from modern spectroscopic surveys, background stars are widely used to show how the integrated columns of DIB carriers 
accumulate from the Sun to great distances. To date, studies on the kinematics of the DIB carriers are still rare.}
{We aim to make use of the measurements from the Giraffe Inner Bulge Survey (GIBS) and the Gaia--ESO survey (GES) to study the kinematics 
and distance of the carrier of DIB\,$\lambda$8620, as well as other properties.}
{The DIBs were detected and measured following the same procedures as in \citetalias{hz21}, assuming a Gaussian profile. The median 
radial velocities of the DIB carriers in 38 GIBS and GES fields were used to trace their kinematics, and the median distances of the 
carriers in each field were estimated by the median radial velocities and two applied Galactic rotation models.}
{We successfully detected and measured DIB\,$\lambda$8620 in 760 of 4117 GES spectra with $|b|\,{\leqslant}\,10^{\circ}$ and 
$\rm S/N>50$. Combined with the DIBs measured in GIBS spectra (\citetalias{hz21}), we confirmed a tight relation between EW and $\EJKs$ 
as well as $\Av$, with similar fitting coefficients to those found by previous works. With a more accurate sample and the consideration 
of the solar motion, the rest-frame wavelength of DIB\,$\lambda$8620 was redetermined as 8620.83\,{\AA}, with a mean fit error 
of 0.36\,{\AA}. We studied the kinematics of the DIB carriers by tracing their median radial velocities in each field in the local 
standard of rest ($\Vlsr$) and into the galactocentric frame ($\Vgc$), respectively, as a function of the Galactic longitudes.
Based on the median $\Vlsr$ and two Galactic rotation models, we obtained valid kinematic distances of the DIB carriers for nine 
GIBS and ten GES fields. We also found a linear relation between the DIB\,$\lambda$8620 measured in this work and the near-infrared 
DIB in APOGEE spectra at $1.5273\,\um$, and we estimated the carrier abundance to be slightly lower compared to the DIB\,$\lambda$15273.}
{We demonstrate that the DIB carriers can be located much closer to the observer than the background stars based on the following arguments: 
(i) qualitatively, the carriers occupy in the Galactic longitude--velocity diagram typical rotation velocities of stars in the local 
Galactic disk, while the background stars in the GIBS survey are mainly located in the Galactic bulge; (ii) quantitatively, all 
the derived kinematic distances of the DIB carriers are smaller than the median distances to background stars in each field. A linear 
correlation between DIB\,$\lambda$8620 and DIB\,$\lambda$15273 has been established, showing similar carrier abundances and making 
them both attractive for future studies of the interstellar environments.}

\maketitle
%
\section{Introduction}

Diffuse interstellar bands (DIBs) are a set of absorption features that can be observed nearly everywhere in the spectra from optical
to infrared wavelengths. DIBs were first observed in 1919 \citep{Heger1922} and then named and definitively determined as interstellar 
features in the 1930s \citep{Merrill38}. Strong optical DIBs, such as the famous $\lambda$5780 and $\lambda$5797 \citep{Heger1922} and 
$\lambda$6284 and $\lambda$6614 \citep{Merrill30}, were discovered and studied first. Then, with the increase of spectral resolution, 
more and more DIBs, especially weak features, were reported. An example mentioned in \citet{Krelowski18}, a recent review, was that 
in the spectral window 5700--5860\,{\AA}, \citet{Heger1922} mentioned two DIBs, \citet{Herbig75} mentioned five, and after, with \citet{Hobbs09}, 
the number increased to 30. In the latest published catalog \citep{Fan19}, this range contained 51 DIBs. Of course, most of them 
are very weak features that can only be detected in high-resolution, high signal-to-noise (S/N) spectra. DIBs were also discovered 
at infrared bands \citep[e.g.,][]{Cox14,Hamano2015,Galazutdinov17a} and in distant galaxies \citep{Monreal-Ibero15,Monreal-Ibero18}. 

The most important topic with regard to the DIBs is the identification of their carriers, which is usually described as ``the longest-standing 
unsolved mystery in astronomical spectroscopy'' (see \citealt{Zack14}, \citealt{Tielens14}, \citealt{Geballe16}, and \citealt{Krelowski18}
for recent reviews). Today, 
carbon-bearing molecules, such as carbon chains \citep{Maier04}, polycyclic aromatic hydrocarbons \citep[PAHs,][]{Omont19}, and fullerenes 
\citep{Omont2016}, are thought to be the most likely candidates for the DIB carriers. As the direct comparison between the observations 
and the laboratory predictions is very difficult, buckminsterfullerene $(C_{60}^{+})$ is the first and only identified DIB carrier 
for five near-infrared (NIR) DIBs \citep{Campbell15,Walker2016,Campbell2016b,Campbell2016a,Cordiner17,Walker2017,CM2018,Lallement18,
Cordiner19,Linnartz2020}, although some debates still exist \citep[e.g.,][]{Galazutdinov17b,Galazutdinov2021}. Besides the laboratory 
investigations, modern spectroscopy also provides observational clues for the carrier research. Mutual correlations between different 
DIBs \citep[e.g.,][]{McCall10,Friedman2011,Elyajouri17} facilitate the estimation of a single carrier for a set of DIBs, although it is still 
too early to conclude the same origin for any pair because of the variation of their strength ratio \citep{Krelowski16}. \citet{Elyajouri18} 
reported a tight correlation between the strength of the so-called $C_2$-DIBs \citep{Thorburn03} and the $C_2$ column density, which 
is very different from other non-$C_2$ DIBs. They also discovered substructures for at least 14 $C_2$-DIBs, which may reveal information 
about the rotational branches of the carriers. It should be noted that although dust grain lost its qualification as the carrier candidate
\citep[see e.g.,][]{Cox07,Cox11,XiangFY17}, the tight correlation between the DIB strength and interstellar extinction for many strong 
DIBs \citep[e.g.,][]{Lan15} means that the DIB carriers are well mixed with the interstellar dust grains. 

The studies related to the DIB carriers focus on their physical identifications, while very few works pay attention to the kinematics 
and distances of the DIB carriers. As the statistical research was not available during the early studies, which usually made use of 
only a few to tens of spectra, the unresolved carriers alleviate the importance and necessity of the kinematic study. However, kinematic 
research benefits from the large spectroscopic surveys and can reveal the rotation curve of the DIB carriers, such as the longitude--velocity 
diagram built for the NIR DIB\,$\lambda$15273 \citep{Zasowski15} based on the APOGEE spectra \citep{Eisenstein11}. The three-dimensional 
(3D) distribution of the DIB carrier is revealed by modern spectroscopic surveys \citep[e.g.,][]{Kos14,Zasowski15}. Furthermore, DIBs 
are also proven to be good tracers of Galactic arms \citep{Puspitarini15,PL2019}. In this work, we aim to establish a kinematic study of 
the DIB\,$\lambda$8620 (although the rest-frame wavelength, $\lambda_0$, for this DIB is larger than 8620\,{\AA}, we still call it 
DIB\,$\lambda$8620 for brevity) and estimate the kinematic distance of its carrier.

The DIB\,$\lambda$8620 was first observed in the spectrum of the star HD 183143 \citep{Geary75} and then confirmed as an interstellar 
band by \citet{Sanner78}, who further reported $\lambda_0\,{=}\,8620.7\,{\pm}\,0.3$\,{\AA} and derived a linear correlation between 
the DIB strength and $\EBV$. Later on, various $\lambda_0$ were measured by different works: 8620.75\,{\AA} \citep{HL1991}, 8621.2\,{\AA} 
\citep{Jenniskens94}, 8620.8\,{\AA} \citep{Galazutdinov00}, and 8620.18\,{\AA} \citep{Fan19}. In the new century, DIB\,$\lambda$8620 
attracts more attention because it is within the spectral window of the RAdial Velocity Experiment (RAVE) survey \citep{Steinmetz06}. 
A tight correlation with $\EBV$ was confirmed by \citet{Munari08} using 68 hot stars from RAVE, and $\lambda_0=8620.4\pm0.1$\,{\AA} 
was suggested. Moreover, \citet{Kos13} and \citet{Kos14} built the first pseudo-3D intensity map for DIB\,$\lambda$8620 using 
$\sim$500,000 RAVE spectra. The correlation between the DIB strength and interstellar extinction was also studied by 
\citet{Wallerstein07}, \citet{Puspitarini15}, and \citet{Damineli16}. Nearly all the linear coefficients derived by different works 
are slightly different from each other, but most of them still correspond with each other considering the uncertainties. 
\citet[][hereafter Paper\,I]{hz21} developed a set of procedures for automatic detection and measurement of the DIB\,$\lambda$8620 
in a spectral window between 8605 and 8640\,{\AA}. The DIB quantities, depth, width, central wavelength ($\lambda_C$), and equivalent 
width (EW), can be measured from the Gaussian profile, together with their uncertainties and the quality flag (QF), which evaluates 
the reliability of the fit. The procedures were tested with 4797 low-resolution spectra from the Giraffe Inner Bulge Survey \citep[GIBS;][]{Zoccali14}, 
a survey of red clump (RC) stars in the Galactic bulge. We derived the EW--$\EJKs$ linear relation in three ways by the 
median quantities in reddening bins, in observational fields, and with a pure RC sample (see Table 1 and Sect 4.1 in \citetalias{hz21} for details). 
The recommended result was $\EJKs\,{=}\,1.884\,({\pm}\,0.225)\,{\times}\,{\rm EW}\,{-}\,0.012\,({\pm}\,0.072)$, which was a medium 
value in comparison with other results and closest to \citet{Munari08}, under the conversion with a specific extinction law. 

In this paper, we continue to study the kinematics and distance of the carrier of DIB\,$\lambda$8620, with high-quality results from 
the GIBS samples used in \citetalias{hz21} and new samples from the {\it Gaia}--ESO Spectroscopic Survey \citep[GES;][]{Gilmore12}. 
In Sect. \ref{sec:data}, we describe the spectra used in this work and define the fields for further analysis. The linear relation 
between EW and extinction is derived and discussed in Sect. \ref{sec:ew-ext}. Section \ref{sec:velocity} presents the kinematic studies 
of the DIB carriers. The estimation of the carrier distance is introduced and discussed in Sect. \ref{sec:distance}. In Sect. 
\ref{sec:apo-compare}, a rough comparison between DIB\,$\lambda$8620 and DIB\,$\lambda$15273 is made to study their possible correlation. 
The main conclusions are summarized in Sect. \ref{sec:conclusion}.

\section{Samples and fields} \label{sec:data}

\begin{figure*}
    \centering
    \includegraphics[width=16cm]{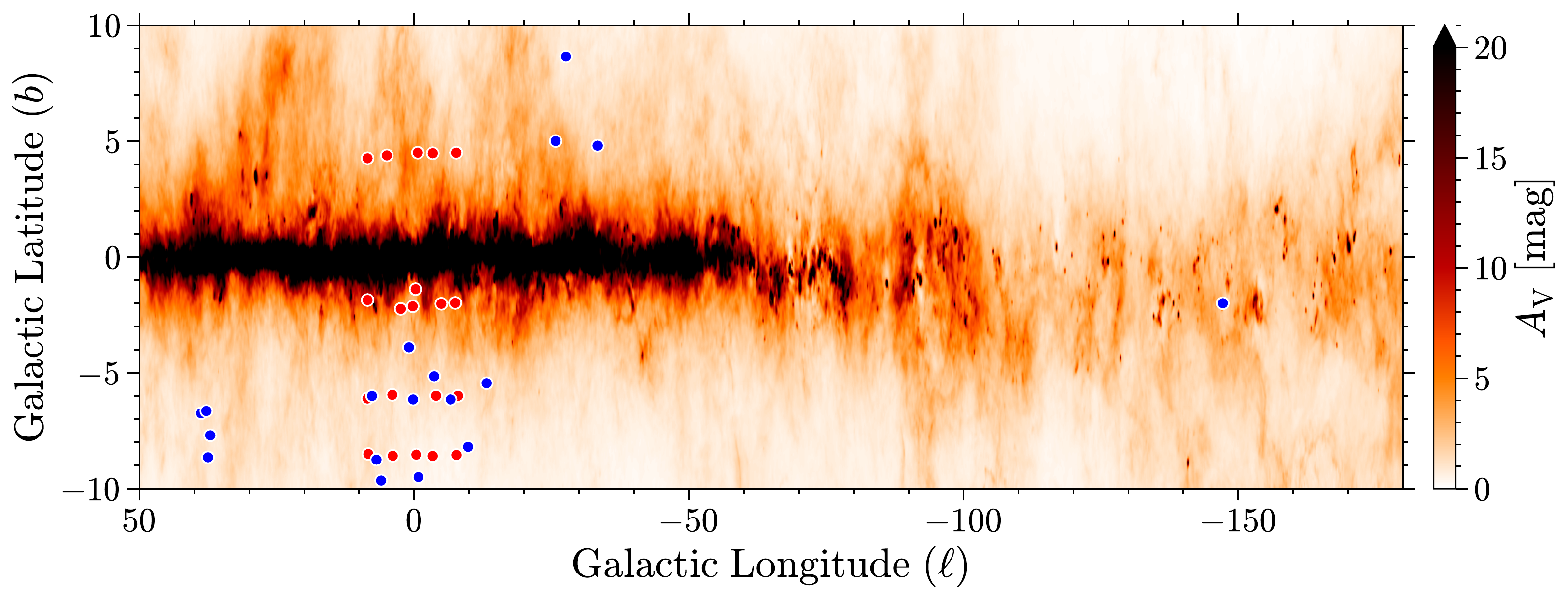}
    \caption{Locations of 20 GIBS observational fields (red dots) and 18 GES selected fields (blue dots), overplotted on the extinction
    map of \citet{SFD} calibrated by \citet{SF11}.}
    \label{fig:field}
\end{figure*}

Giraffe Inner Bulge Survey (GIBS) is a dedicated survey to study the kinematics and chemistry of RC stars in the Galactic bulge
\citep{Zoccali14}. In \citetalias{hz21}, we constructed a pure RC sample applying an additional criterion where we consider stars 
only to be on the RC if their $\JKs$ colors lie within 1$\sigma$ width of the peak of the RC $\JKs$ (see \citetalias{hz21}, Appendix 
A for more details). This guarantees, as discussed in \citetalias{hz21}, a pure RC sample, and avoids contamination by foreground 
dwarfs and/or red giant branch (RGB) stars. In addition, we applied a cut of $\rm S/N\,{>}\,50$ to ensure high-quality measurements. 
Finally, our working sample consists of 1780 DIBs distributed in 20 GIBS fields in total. The DIB measurements of the first ten GIBS 
targets are shown in Table \ref{tab:ges-dib}.

Gaia--ESO (GES) is a public spectroscopic survey targeting all the major components of the Milky Way with the purpose of characterizing the 
chemistry and the kinematics of these populations. A detailed description of the data processing and general characterization of 
the data set can be found in \citet{Gilmore12}. For this paper, we used the official public data release DR4\footnote{\url{https://www.gaia-eso.eu}} 
and the high-resolution grating HR21 centered at $\rm 8757$\,{\AA} with a spectral resolution of $R\,{\sim}\,16,200$ on the GIRAFFE 
spectrograph. We restricted our sample within $|b|\,{\leqslant}\,10^{\circ}$ and $\rm S/N\,{>}\,50,$ which gives a total of 4117
spectra.

The stellar parameters of the GES stars were estimated by applying the MATISSE \citep{Recio-Blanco2006,Recio-Blanco16} parameterization 
algorithm to the corresponding spectra, refining the results thanks to the GAUGUIN procedure \citep{Bijaoui2012,Recio-Blanco16}. On 
one hand, MATISSE is a projection method for which the full input spectra are projected into a set of vectors derived during a learning 
phase, based on the noise-free reference grids. These vectors are a linear combination of reference spectra and could be viewed roughly 
as the derivatives of these spectra with respect to the different stellar parameters. MATISSE is thus a local multi-linear regression 
method. On the other hand, GAUGUIN is a classical local optimization method implementing a Gauss--Newton algorithm. It is based on a 
local linearization around a given set of parameters that are associated with a reference synthetic spectrum (via linear interpolation 
of the derivatives). A few iterations are carried out through linearization around the new solutions, until the algorithm converges 
toward the minimum distance. In this application, GAUGUIN is initialized by the MATISSE parameters solution.

Both parameterization algorithms together with the DIB measurement rely on a grid of synthetic spectra specifically computed for 
FGKM-type stars analyzed by GES. This grid contains high-resolution synthetic spectra over the spectral range 845--895\,nm. It covers 
metallicities from $\rm [M/H]\,{=}\,{-}5.0$ to ${+}1.0$\,dex and variations in $\rm [\alpha/Fe]$ (five values for each metallicity). The 
grid computation adopted the same methodology as the grid computed for the AMBRE project \citep{de-Laverny2013} and is described in 
\citet{de-Laverny2012}. We remind the reader that it is based on the MARCS model atmospheres \citep{Gustafsson08} and the Turbospectrum code 
for radiative transfer \citep{Plez2012}. For the present application, we adopted the GES atomic and molecular line lists \citep{Heiter2021}
and a microturbulence velocity that varies with the atmospheric parameter values (empirical relation adopted within GES, \citealt{Bergemann2021pre}).

\begin{figure}
    \centering
    \includegraphics[width=8cm]{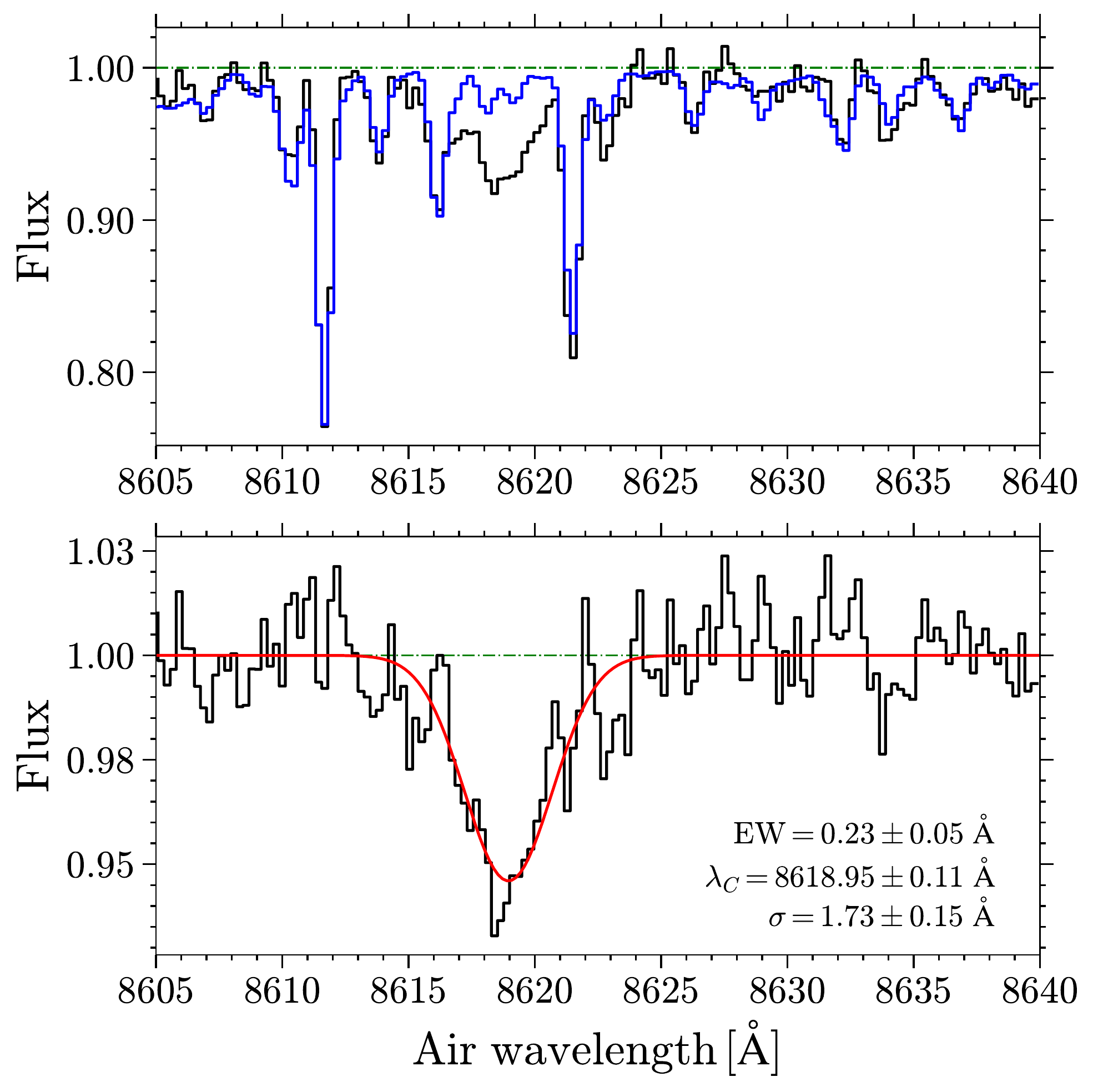}
    \caption{Fit of the DIB\,$\lambda$8620 in the spectrum of the GES target 06440722--0055038. {\it Upper panel:} Black and blue lines 
    show the observed and synthetic spectra, respectively. {\it Lower panel:} Black line is the renormalized interstellar spectrum. 
    The red line represents the fitted Gaussian profile. The measured EW, central wavelength ($\lambda_C$), and width ($\sigma$)
    are also indicated.}
    \label{fig:fit-example}
\end{figure}

The DIBs in GES spectra were detected and measured by the procedures developed in \citetalias{hz21}. We were able to recover l 760 
DIBs in tota with $\rm QF\,{>}\,0$. The fit results for the first ten GES targets are shown in Table \ref{tab:ges-dib}, and the full catalog 
can be accessed online. A fit example is shown in Fig. \ref{fig:fit-example}. Due to the DIBs' sparse 
sampling, we manually defined 18 GES fields with at least ten stars in each field (see Fig. \ref{fig:field}).

In total, our combined GIBS and GES sample consists of 20 GIBS fields and 18 GES fields. Figure \ref{fig:field} shows the spatial 
distribution of the fields, overplotted with the extinction map of \citet[][hereafter SFD]{SFD}, calibrated by \citet{SF11}. Table 
\ref{tab:field} lists the central coordinates and radii of each field, together with the number of DIBs in them. 

\begin{table*}
\begin{center}
\small
\caption{Fit results of DIB\,$\lambda$8620 in the GIBS and GES data sets, as well as extinction, distance, and radial
         velocity of the background stars. \label{tab:ges-dib}}
\begin{tabular}{l c c c c c c c c r}
\hline\hline 
target ID & source & $\ell$\tablefootmark{1} & $b$        & $\lambda_C \pm {\rm err}$\tablefootmark{2} & $\sigma \pm {\rm err}$\tablefootmark{3} 
                   & $\rm EW \pm {\rm err}$\tablefootmark{4} & $\EJKs$\tablefootmark{5}  & $d_{\rm star}$\tablefootmark{6} & $V_{\rm rad}$\tablefootmark{7} \\
          &        & $(^\circ)$              & $(^\circ)$ & ({\AA})                                    & ({\AA})  
                   & ({\AA})                                 & (mag)                     & (kpc)                           & $(\rm km\,s^{-1})$ \\ [0.5ex]
\hline 
06432248--0046204 & GES & --147.30 & --2.15 & $8622.46\,{\pm}\,0.02$ & $0.80\,{\pm}\,0.03$ & $0.063\,{\pm}\,0.026$ & 0.21 & 2.45 &   --5.62 \\
06432400--0046576 &     GES & --147.29 & --2.15 & $8616.15\,{\pm}\,0.07$ & $1.89\,{\pm}\,0.12$ & $0.107\,{\pm}\,0.023$ & 0.19 & 3.24 &   171.91 \\
06432685--0047112 &     GES & --147.28 & --2.14 & $8620.20\,{\pm}\,0.25$ & $1.80\,{\pm}\,0.23$ & $0.120\,{\pm}\,0.008$ & 0.22 & 3.09 &   54.54  \\
06433215--0055203 &     GES & --147.15 & --2.18 & $8621.57\,{\pm}\,0.03$ & $0.72\,{\pm}\,0.04$ & $0.149\,{\pm}\,0.045$ & 0.26 & 0.55 &   11.40  \\
06433789--0047286 &     GES & --147.25 & --2.10 & $8620.37\,{\pm}\,0.24$ & $1.85\,{\pm}\,0.12$ & $0.146\,{\pm}\,0.012$ & 0.25 & 5.22 &   64.12  \\
06434241--0052100 &     GES & --147.18 & --2.12 & $8621.57\,{\pm}\,0.02$ & $0.78\,{\pm}\,0.04$ & $0.069\,{\pm}\,0.021$ & 0.13 & 1.11 &   12.63  \\
06434513--0103076 &     GES & --147.01 & --2.19 & $8619.63\,{\pm}\,0.08$ & $1.43\,{\pm}\,0.07$ & $0.109\,{\pm}\,0.028$ & 0.25 & 4.33 &   69.55  \\
06434800--0051432 &     GES & --147.17 & --2.09 & $8619.81\,{\pm}\,0.20$ & $2.11\,{\pm}\,0.29$ & $0.194\,{\pm}\,0.018$ & 0.35 & 5.58 &   65.63  \\
06435592--0057034 &     GES & --147.08 & --2.10 & $8618.98\,{\pm}\,0.03$ & $0.74\,{\pm}\,0.07$ & $0.052\,{\pm}\,0.007$ & 0.21 & 2.08 &   97.97  \\
06435624--0040198 &     GES & --147.32 & --1.98 & $8621.36\,{\pm}\,0.06$ & $1.66\,{\pm}\,0.05$ & $0.135\,{\pm}\,0.010$ & 0.18 & 1.50 &   36.71  \\
$\cdots$          &     &          &        &                        &                     &                       &      &      &          \\
LRp8p4\_F1\_4094  & GIBS &  8.32   &   4.31 & $8622.01\,{\pm}\,0.36$ & $1.74\,{\pm}\,0.38$ & $0.169\,{\pm}\,0.050$ & 0.46 & 7.56 &  --38.34 \\ 
LRp8p4\_F1\_4343  &     GIBS &  8.34   &   4.39 & $8623.98\,{\pm}\,0.46$ & $1.88\,{\pm}\,0.49$ & $0.102\,{\pm}\,0.034$ & 0.48 & 7.24 &   --7.38 \\ 
LRp8p4\_F1\_4355  &     GIBS &  8.36   &   4.36 & $8622.18\,{\pm}\,0.28$ & $1.42\,{\pm}\,0.34$ & $0.146\,{\pm}\,0.006$ & 0.52 & 7.29 &  --36.61 \\ 
LRp8p4\_F1\_4379  &     GIBS &  8.38   &   4.32 & $8624.13\,{\pm}\,0.38$ & $1.70\,{\pm}\,0.43$ & $0.117\,{\pm}\,0.015$ & 0.40 & 7.80 & --137.54 \\ 
LRp8p4\_F1\_4393  &     GIBS &  8.40   &   4.35 & $8620.88\,{\pm}\,0.45$ & $1.81\,{\pm}\,0.45$ & $0.129\,{\pm}\,0.028$ & 0.44 & 7.81 &   --7.85 \\ 
LRp8p4\_F1\_4399  &     GIBS &  8.41   &   4.39 & $8614.71\,{\pm}\,0.40$ & $1.53\,{\pm}\,0.41$ & $0.145\,{\pm}\,0.044$ & 0.49 & 7.76 &   182.51 \\ 
LRp8p4\_F1\_4400  &     GIBS &  8.41   &   4.35 & $8622.18\,{\pm}\,0.29$ & $1.59\,{\pm}\,0.40$ & $0.243\,{\pm}\,0.022$ & 0.44 & 7.25 &  --22.18 \\ 
LRp8p4\_F1\_4422  &     GIBS &  8.33   &   4.28 & $8614.35\,{\pm}\,0.41$ & $1.52\,{\pm}\,0.42$ & $0.109\,{\pm}\,0.017$ & 0.43 & 8.07 &   189.82 \\ 
LRp8p4\_F1\_4424  &     GIBS &  8.34   &   4.28 & $8623.01\,{\pm}\,0.47$ & $2.00\,{\pm}\,0.44$ & $0.186\,{\pm}\,0.022$ & 0.41 & 7.75 &    13.02 \\ 
LRp8p4\_F1\_4444  &     GIBS &  8.37   &   4.25 & $8618.93\,{\pm}\,0.39$ & $1.68\,{\pm}\,0.56$ & $0.118\,{\pm}\,0.018$ & 0.41 & 7.22 &    68.52 \\ 
$\cdots$          &     &          &        &                        &                     &                       &      &      &          \\ [0.5ex]
\hline
\end{tabular}
\end{center}
\tablefoot{Full table can be accessed online.\\
\tablefoottext{1}{$\ell \in \pm180^{\circ}$} \\
\tablefoottext{2}{Measured central wavelength in the stellar frame.} \\
\tablefoottext{3}{The width of the DIB profile.} \\
\tablefoottext{4}{Equivalent width.} \\
\tablefoottext{5}{$\EJKs$ is from \citetalias{Alvaro17} for GES targets and from \citetalias{Surot2020} for GIBS targets.} \\
\tablefoottext{6}{Distances to the background stars; see Sect. \ref{subsec:star-dist} for details.} \\
\tablefoottext{7}{Stellar radial velocity in the heliocentric frame.}}
\end{table*}

\begin{table*}
\begin{center}
\caption{Selected GIBS and GES fields. \label{tab:field}}
\begin{tabular}{l c c c r c c c c}
\hline\hline 
Field    & ($\ell_0$,\,$b_0$) & radius     & source & DIB     & $d_{\rm stellar} \pm {\rm err}$\tablefootmark{1} 
         & $R_{\rm GC}^{\rm Reid19} \pm {\rm err}$\tablefootmark{2} & $R_{\rm GC}^{\rm Mroz19} \pm {\rm err}$\tablefootmark{3}
         & $d_{\rm los} \pm {\rm err}$\tablefootmark{4} \\ 
Nr       & $(^\circ)$         & $(^\circ)$ &        & Nr      & (kpc)  & (kpc) & (kpc) & (kpc)  \\ [0.5ex]
\hline 
1     & (8.5,\,4.3)       & 0.2 & GIBS & 123 & $7.68\,{\pm}\,0.41$ & $4.69\,{\pm}\,0.45$        & $4.87\,{\pm}\,0.36$ & $3.57\,{\pm}\,0.48$  \\
2     & (5.0,\,4.4)       & 0.2 & GIBS & 116 & $7.30\,{\pm}\,0.37$ & --                     & --                  & --                   \\
3     & (--0.7,\,4.5)     & 0.2 & GIBS & 116 & $9.04\,{\pm}\,0.47$ & $6.99\,{\pm}\,4.59$        & $6.97\,{\pm}\,1.74$ & --                   \\
4     & (--3.4,\,4.5)     & 0.1 & GIBS & 107 & $8.58\,{\pm}\,0.44$ & $5.06\,{\pm}\,1.30$        & $5.33\,{\pm}\,1.12$ & $3.17\,{\pm}\,1.19$  \\
5     & (--7.7,\,4.5)     & 0.2 & GIBS & 101 & $9.23\,{\pm}\,0.50$ & $7.65\,{\pm}\,1.04$        & $7.67\,{\pm}\,0.88$ & --                   \\
6     & (--0.3,\,--1.4)   & 0.2 & GIBS & 205 & $7.44\,{\pm}\,0.91$ & --                     & --                  & --                   \\
7     & (8.5,\,--1.9)     & 0.1 & GIBS & 54  & $7.45\,{\pm}\,0.55$ & $6.48\,{\pm}\,0.47$        & $6.44\,{\pm}\,0.39$ & $1.80\,{\pm}\,0.48$  \\
8     & (2.4,\,--2.2)     & 0.1 & GIBS & 99  & $7.80\,{\pm}\,0.46$ & $4.18\,{\pm}\,0.98$        & $4.67\,{\pm}\,0.86$ & $3.86\,{\pm}\,0.99$  \\
9     & (0.3,\,--2.1)     & 0.1 & GIBS & 171 & $7.58\,{\pm}\,0.81$ & --                     & --                  & --                   \\
10    & (--4.9,\,--2.0)   & 0.1 & GIBS & 128 & $9.05\,{\pm}\,0.33$ & $6.80\,{\pm}\,0.66$        & $6.64\,{\pm}\,0.52$ & $1.50\,{\pm}\,0.62$  \\
11    & (--7.5,\,--2.0)   & 0.2 & GIBS & 105 & $9.16\,{\pm}\,0.46$ & $7.17\,{\pm}\,0.57$        & $7.01\,{\pm}\,0.45$ &   $0.93\,{\pm}\,0.52$  \\
12    & (--8.5,\,--6.1)   & 0.2 & GIBS & 63  & $7.49\,{\pm}\,0.43$ & $4.98\,{\pm}\,1.04$        & $5.18\,{\pm}\,0.95$ &   $3.22\,{\pm}\,1.03$  \\
13    & (4.0,\,--6.0)     & 0.2 & GIBS & 48  & $7.54\,{\pm}\,1.10$ & --                          & --                  & --                   \\
14    & (--4.0,\,--6.0)   & 0.2 & GIBS & 88  & $9.24\,{\pm}\,0.65$ & $7.86\,{\pm}\,3.13$        & $7.63\,{\pm}\,1.36$ &   --                   \\
15    & (--8.0,\,--6.0)   & 0.2 & GIBS & 55  & $9.45\,{\pm}\,0.91$ & $11.48\,{\pm}\,3.97$       & $8.82\,{\pm}\,0.97$ &   --                   \\
16    & (8.3,\,--8.5)     & 0.2 & GIBS & 54  & $7.44\,{\pm}\,0.47$ & $4.56\,{\pm}\,0.78$        & $4.81\,{\pm}\,0.66$ &   $3.72\,{\pm}\,0.81$  \\
17    & (3.9,\,--8.6)     & 0.2 & GIBS & 55  & $7.14\,{\pm}\,0.38$ & --                     &  --                 &      --                   \\
18    & (--0.4,\,--8.5)   & 0.2 & GIBS & 52  & $7.47\,{\pm}\,0.35$ & --                     &  --                 &      --                   \\
19    & (--3.4,\,--8.6)   & 0.2 & GIBS & 20  & $9.91\,{\pm}\,0.46$ & $5.25\,{\pm}\,4.37$    & $5.80\,{\pm}\,2.03$ &      --                   \\
20    & (--7.7,\,--8.5)   & 0.2 & GIBS & 20  & $9.29\,{\pm}\,0.62$ & $4.60\,{\pm}\,2.55$        & $5.34\,{\pm}\,1.73$ &   $3.49\,{\pm}\,1.78$  \\
21    & (--147.2,\,--2.0) & 0.2 & GES  & 22  & $2.84\,{\pm}\,1.47$ & $9.95\,{\pm}\,0.33$        & $9.73\,{\pm}\,0.17$ & $2.07\,{\pm}\,0.37$  \\
22    & (--33.4,\,4.8)    & 0.2 & GES  & 91  & $8.80\,{\pm}\,1.67$ & $6.80\,{\pm}\,0.20$        & $6.66\,{\pm}\,0.06$ & $1.76\,{\pm}\,0.27$  \\
23    & (--25.8,\,5.0)    & 0.2 & GES  & 48  & $6.10\,{\pm}\,1.70$ & $7.53\,{\pm}\,0.26$        & $7.38\,{\pm}\,0.13$ & $0.74\,{\pm}\,0.29$  \\
24    & (--27.7,\,8.7)    & 0.2 & GES  & 19  & $5.84\,{\pm}\,2.09$ & $8.35\,{\pm}\,0.43$        & $8.20\,{\pm}\,0.36$ & --                   \\
25    & (--13.2,\,--5.5)  & 0.2 & GES  & 28  & $5.63\,{\pm}\,2.80$ & $8.52\,{\pm}\,0.67$        & $8.38\,{\pm}\,0.57$ & --                   \\
26    & (--9.8,\,--8.2)   & 0.4 & GES  & 44  & $5.21\,{\pm}\,1.98$ & $9.67\,{\pm}\,1.47$        & $9.08\,{\pm}\,0.68$ & --                   \\
27    & (--6.7,\,--6.2)   & 0.3 & GES  & 24  & $5.24\,{\pm}\,1.81$ & $6.41\,{\pm}\,1.37$        & $6.58\,{\pm}\,1.14$ & $1.80\,{\pm}\,1.02$  \\
28    & (--3.7,\,--5.2)   & 0.2 & GES  & 16  & $4.77\,{\pm}\,2.11$ & --                     & --                  & --                   \\
29    & (1.0,\,--3.9)     & 0.3 & GES  & 91  & $4.86\,{\pm}\,2.13$ & --                     & --                  & --                   \\
30    & (0.2,\,--6.2)     & 0.2 & GES  & 12  & $6.78\,{\pm}\,2.05$ & --                     & --                  & --                   \\
31    & (--0.8,\,--9.5)   & 0.2 & GES  & 13  & $4.06\,{\pm}\,1.52$ & --                     & --                  & --                   \\
32    & (6.0,\,--9.7)     & 0.3 & GES  & 40  & $3.18\,{\pm}\,1.97$ & $8.38\,{\pm}\,1.53$        & $8.19\,{\pm}\,0.97$ & --                   \\
33    & (6.8,\,--8.8)     & 0.4 & GES  & 24  & $5.02\,{\pm}\,1.74$ & $4.90\,{\pm}\,0.88$        & $5.17\,{\pm}\,0.78$ & $3.47\,{\pm}\,0.91$  \\
34    & (7.7,\,--6.0)     & 0.2 & GES  & 27  & $4.62\,{\pm}\,1.14$ & $6.69\,{\pm}\,0.90$        & $6.81\,{\pm}\,0.84$ & $1.43\,{\pm}\,0.77$  \\
35    & (37.5,\,--8.7)    & 0.3 & GES  & 41  & $5.04\,{\pm}\,2.06$ & $7.64\,{\pm}\,0.32$        & $7.50\,{\pm}\,0.24$ & $0.65\,{\pm}\,0.39$  \\
36    & (38.8,\,--6.8)    & 0.3 & GES  & 25  & $3.27\,{\pm}\,2.16$ & $6.76\,{\pm}\,0.38$        & $6.70\,{\pm}\,0.31$ & $1.88\,{\pm}\,0.57$  \\
37    & (37.8,\,--6.7)    & 0.8 & GES  & 112 & $3.18\,{\pm}\,2.29$ & $7.15\,{\pm}\,0.24$        & $7.04\,{\pm}\,0.14$ & $1.34\,{\pm}\,0.34$  \\
38    & (37.1,\,--7.7)    & 0.5 & GES  & 83  & $2.99\,{\pm}\,1.99$ & $7.00\,{\pm}\,0.27$        & $6.92\,{\pm}\,0.18$ & $1.50\,{\pm}\,0.38$  \\ [0.5ex]
\hline
\end{tabular}
\end{center}
\tablefoot{ \\
\tablefoottext{1}{Median stellar distance in each field.} \\
\tablefoottext{2}{Galactocentric distance of the DIB carrier, calculated by the field-median $\Vlsr$ and \citet{Reid19} rotation model.} \\
\tablefoottext{3}{Galactocentric distance of the DIB carrier, calculated by the field-median $\Vlsr$ and \citet{Mroz19} rotation model.} \\
\tablefoottext{4}{Line-of-sight distance of the DIB carrier, derived by $R_{\rm GC}^{\rm Reid19}$, field-median $\ell$, and $\Rsun=8.15$\,kpc.}
}
\end{table*}

\section{Equivalent width and extinction} \label{sec:ew-ext}

The tight linear correlation between EW and interstellar extinction for DIB\,$\lambda$8620 has been reported in many works
\citep[e.g.,][]{Wallerstein07,Munari08,Kos13,Puspitarini15} and also plays an important role in the study of the property of the DIB 
carrier. In \citetalias{hz21}, we derived a linear relation of $\EJKs\,{=}\,1.884\,({\pm}\,0.225)\,{\times}\,{\rm EW}\,{-}\,0.012\,({\pm}\,0.072)$ 
with a pure RC sample, where EW was the median value in each GIBS field and $\EJKs$ in each field was derived based on the 
peak color estimated by the VVV--DR2 catalog \citep{Minniti17} and intrinsic color given by \citet{Gonzalez11}. In this section, 
we briefly describe our measurements of the EW and the $\EJKs$ for the GES targets, while we use the results from 
\citetalias{hz21} for GIBS targets.

\subsection{EW measurement} \label{subsec:ew}

\begin{figure}
    \centering
    \includegraphics[width=8.4cm]{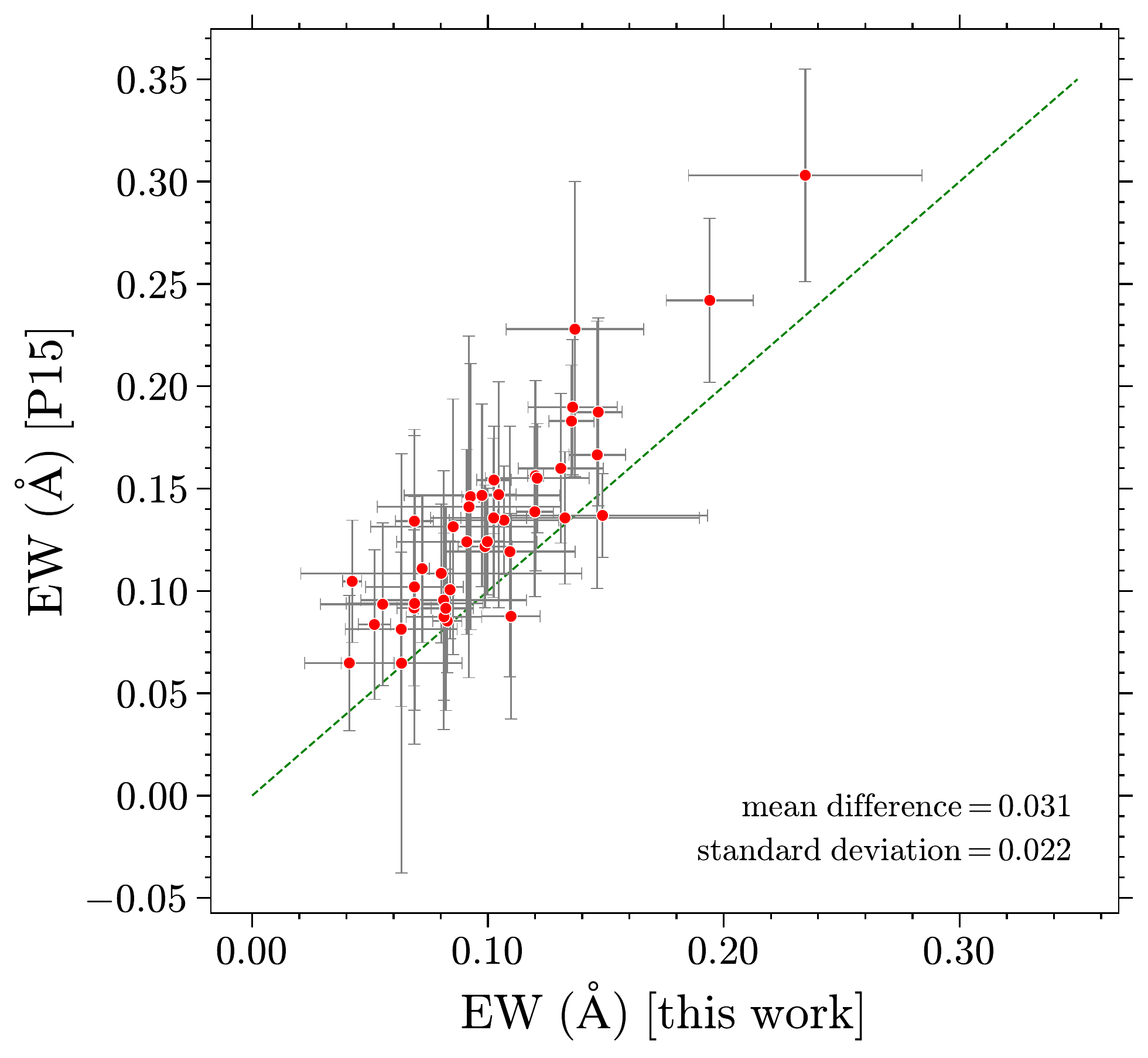}
    \caption{Comparison of the EW computed in this work with the one from \citet{Puspitarini15} for 43 common targets.
    The dashed green line traces the one-to-one correspondence.}
    \label{fig:P15}
\end{figure}

As the DIB profile is fitted by a Gaussian function in our procedures, the EW can be calculated by the fitted depth ($D$) and width 
($\sigma$), $\rm EW\,{=}\,\sqrt{2 \pi}\,{\it D}\,\sigma$. The error of EW is estimated using the same method as described in \citetalias{hz21}, 
considering the contribution of both the random noise (i.e., S/N) and the error based on the discrepancy between the observed and the 
synthetic spectrum. \citet{Puspitarini15} also analyzed the DIB\,$\lambda$8620 in 162 GES spectra. Due to the low S/N of these spectra, 
only 43 passed our quality-flag criteria $(\rm QF\,{>}\,0)$. Figure \ref{fig:P15} shows the comparison of the EW for these 43 stars 
where \citet{Puspitarini15} systematically obtained slightly larger EW than the result in this work, with a mean difference of 0.031\,{\AA} 
and a standard deviation of 0.022\,{\AA}. The mean difference is similar to the average error of EW in this work (0.020\,{\AA}) and 
in that of \citet[][0.045\,{\AA}]{Puspitarini15}. The systematic difference might be caused by the use of different synthetic spectra 
in \citet{Puspitarini15} and this work. Specifically, the synthetic model used in \citet{Puspitarini15} was based on an the ATLAS 9 model 
atmosphere and the SYNTHE suite \citep{Kurucz2005,Sbordone2004,Sbordone2005}, which is different from the MARCS model atmospheres 
\citep{Gustafsson08} used in this work. The different fit methods also contribute to the discrepancy in EW, that is we applied 
a Gaussian fit to the DIB profile, while \citet{Puspitarini15} fitted the DIB feature with an empirical model averaging the profiles 
detected in several spectra based on the data analysis reported by \citet{ChenHC13}.

\subsection{Extinction} \label{subsec:ext}

\begin{figure}
    \centering
    \includegraphics[width=8.4cm]{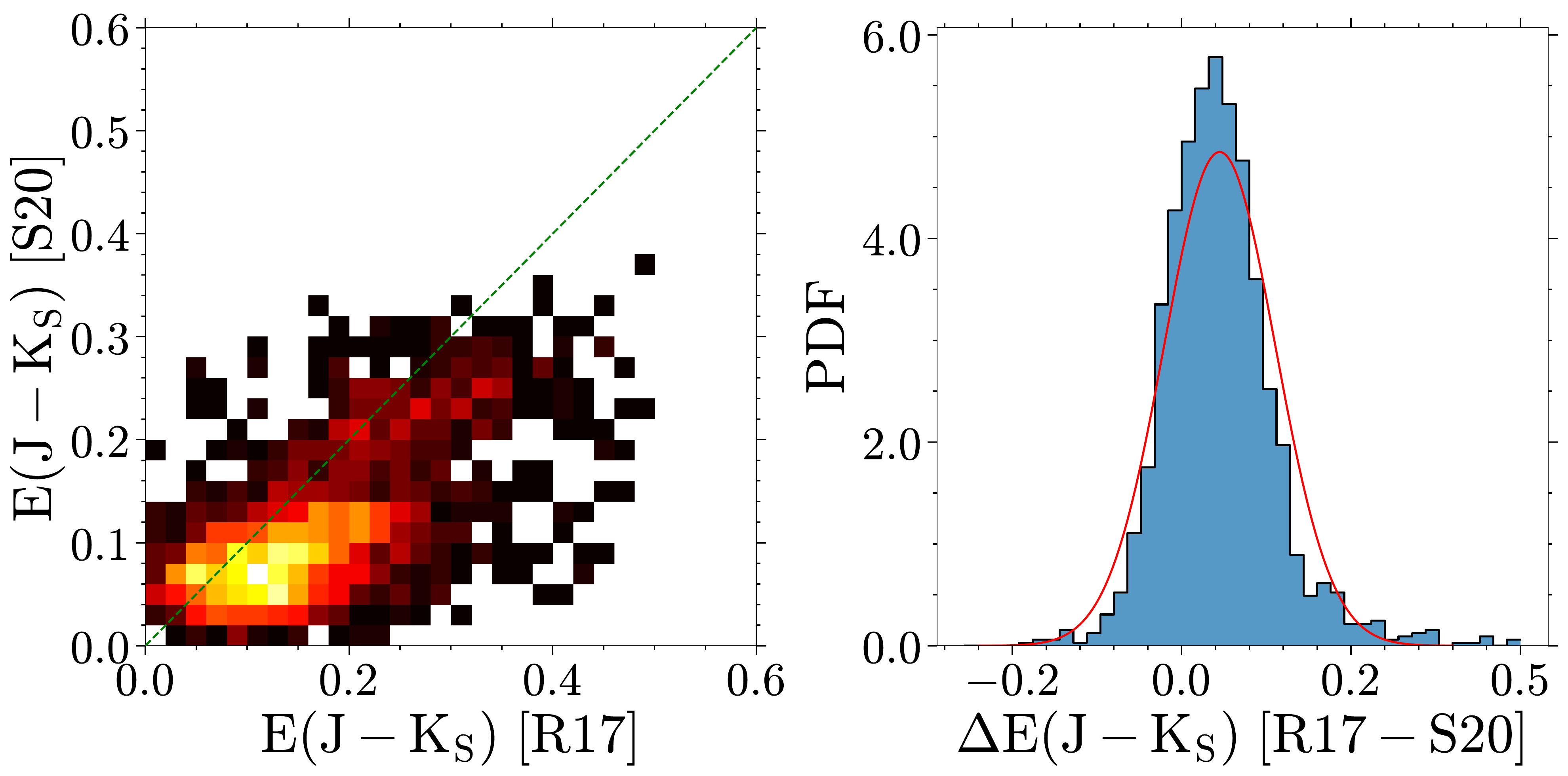}
    \caption{{\it Left panel}: Comparison between $\EJKs$ derived by \citetalias{Alvaro17} and \citetalias{Surot2020}. The dashed 
    green line traces the one-to-one correspondence. The color represents the number density. {\it Right panel}: Distribution of the 
    differences between \citetalias{Alvaro17} and \citetalias{Surot2020}. The red line represents a Gaussian fit.}
    \label{fig:ext-cali}
\end{figure}

The distances to and the individual reddenings of the GES stars were calculated by the spectro-photometric method described in 
\citet[][hereafter R17]{Alvaro17}, using the stellar parameters ($\Teff$, $\logg$, $\feh$) with the corresponding errors together 
with the PARSEC isochrones \citep{Marigo17}. The extinction of our GIBS sample was derived by the high-resolution extinction map 
from \citet[][hereafter S20]{Surot2020} using RGB+RC stars from the VVV survey \citep{Minniti10}. We refer the reader to \citetalias{hz21} 
for a more detailed description. Figure \ref{fig:ext-cali} shows the comparison between $\EJKs$ calculated by \citetalias{Alvaro17} 
and \citetalias{Surot2020} for 1626 GES sample stars with $|\ell|\,{\leqslant}\,10^{\circ}$. The derived $\EJKs_{\rm R17}$ is 
systematically larger than that of $\EJKs_{\rm S20}$ with a mean difference $\rm (R17-S20)$ of 0.056\,mag and a standard deviation 
of 0.082\,mag. As this difference between \citetalias{Alvaro17} and \citetalias{Surot2020} is smaller than the variation of 
$\EJKs$ in each field, we did not attempt to correct for this.

\begin{figure*}
    \centering
    \includegraphics[width=8cm]{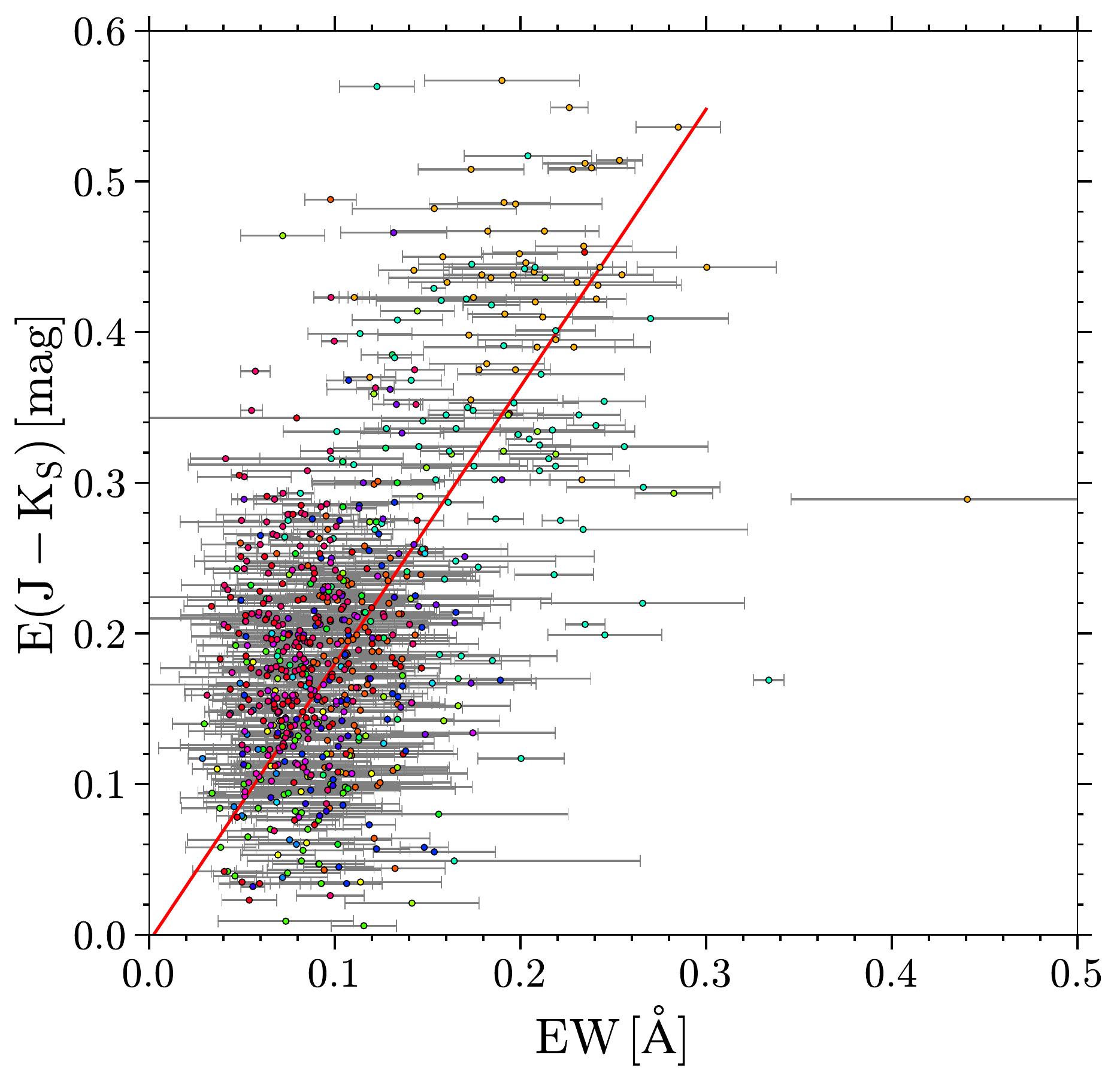}
    \includegraphics[width=8cm]{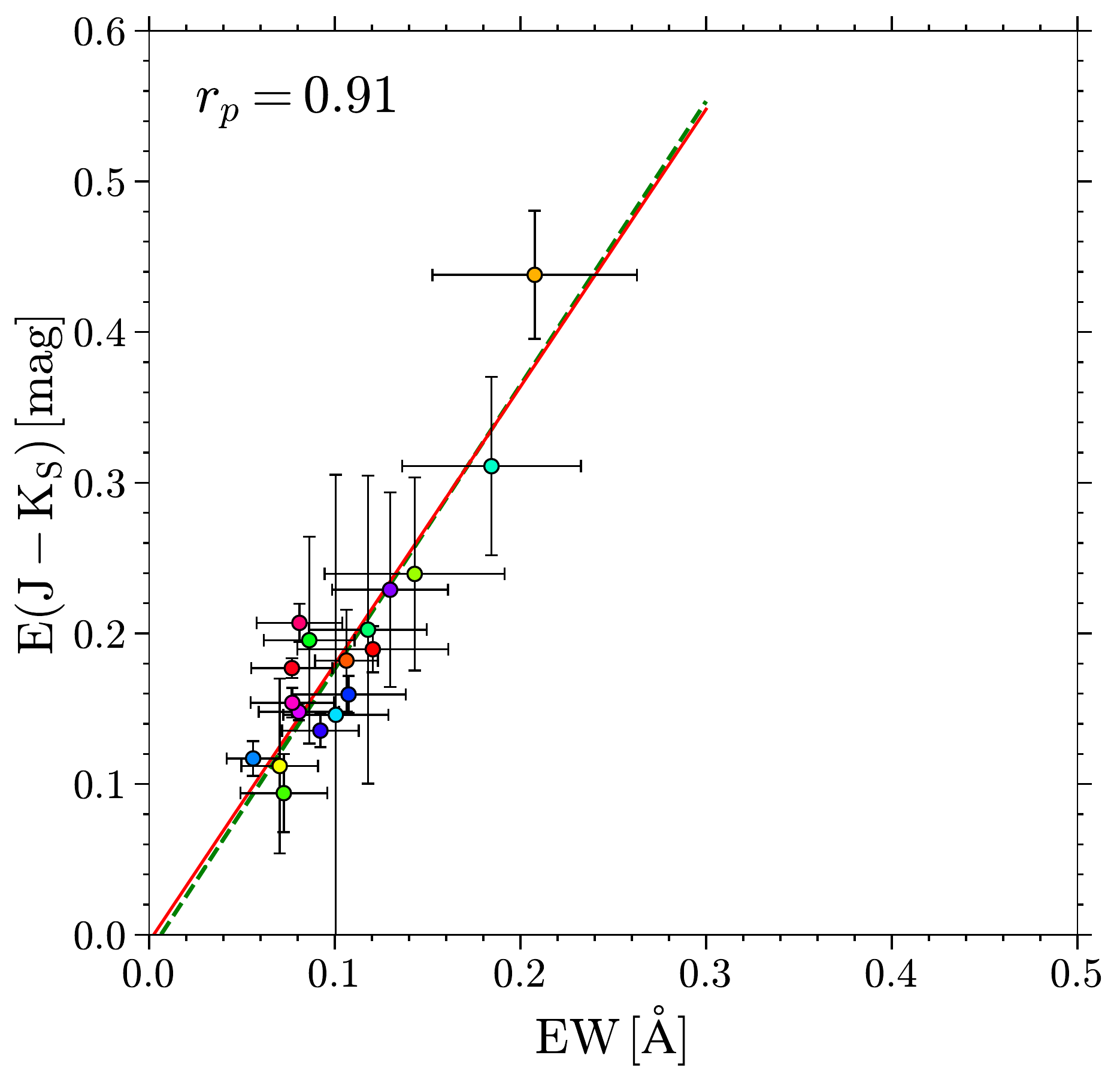}
    \caption{Correlation between EW and $\EJKs$ derived from GES individual targets ({\it left}) and fields ({\it right}). 
    The dots with the same color are from the same field, and their median value shown in right panel is colored in the same way.
    The dashed green line in the right panel shows the relation derived by \citetalias{hz21}. The error bars indicate the standard 
    deviation in each field. The red lines in both panels are fit to the dots in the right panel. The Pearson correlation 
    coefficient ($r_p$) is also indicated.}
    \label{fig:ges-ee}
\end{figure*}

\begin{figure}
    \centering
    \includegraphics[width=8cm]{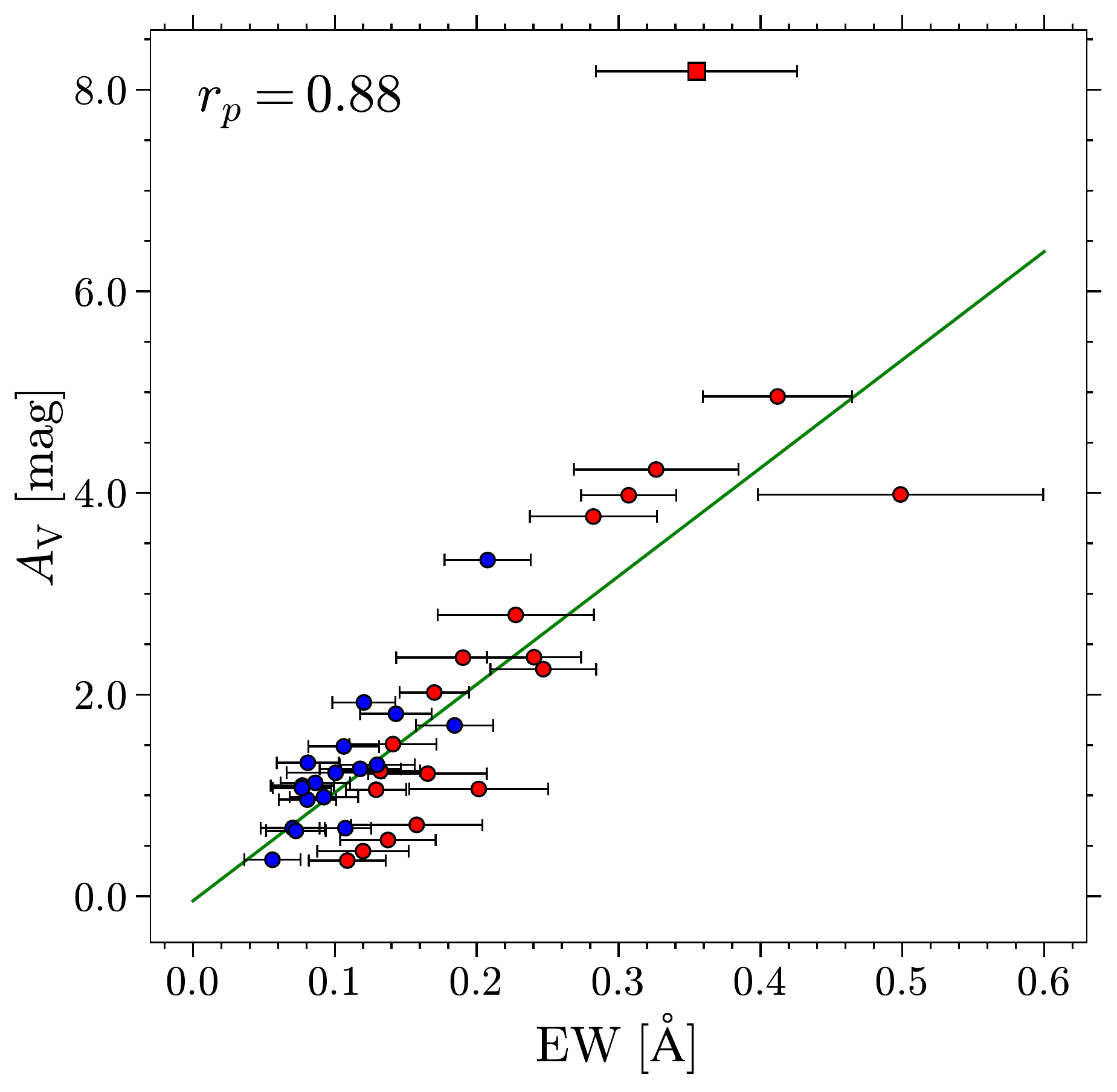}
    \caption{Correlation between EW and $\Av$ \citepalias{SFD} for the median values in each GIBS (red) and GES (blue) field. 
    The green line represents the linear fit, excluding the field with $\Av\,{>}\,8$\,mag (the red square).
    The Pearson correlation coefficient ($r_p$) is also indicated.}
    \label{fig:EW-SFD}
\end{figure}

\subsection{Correlation between EW and extinction} \label{subsec:linear}

Figure \ref{fig:ges-ee} shows the correlation between EW and $\EJKs$ for the individual GES targets (left panel) and the median 
values for each field (right panel). Although a large dispersion is found for individual measurements, the median EW and $\EJKs$ 
for each field present a tight correlation with a Pearson correlation coefficient ($r_p$) of 0.91. The linear relation is derived as
$\EJKs\,{=}\,1.842\,({\pm}\,0.203)\,{\times}\,{\rm EW}\,{-}\,0.005\,({\pm}\,0.023)$, which is highly consistent with the relation 
recommended by \citetalias{hz21} for the GIBS fields (the dashed green line in Fig. \ref{fig:ges-ee}), even when taking into account the 
different resolutions between the GES ($R\,{\sim}\,16,200$) and GIBS ($R\,{=}\,6500$) spectra and the difference between 
$\EJKs_{\rm R17}$ and $\EJKs_{\rm S20}$. For the latter, the reasons are 1) the recommended relation in \citetalias{hz21} was derived 
using an RC-based $\EJKs$ (which is not from the \citetalias{Surot2020} map), although $\EJKs_{\rm S20}$ was used for GIBS individual 
targets. We emphasize here again that in \citetalias{hz21} we derived the relation in three ways, two of them were based on the median 
$\EJKs_{\rm S20}$ in each field, but the preferred one used the RC-based $\EJKs$; and 2) we applied a cut of $\EJKs\,{>}\,0.25$\,mag as 
described in \citetalias{hz21}, and the typical $\EJKs_{\rm R17}$ of GES fields are within 0.3\,mag as they are located at higher 
Galactic latitudes (see Fig. \ref{fig:field}). Thus, the two relations were derived in different $\EJKs$ ranges. Their consistency
might indicate a decrease of the difference between \citetalias{Alvaro17} and \citetalias{Surot2020} for higher $\EJKs$. For the 
further analysis, we use as in \citetalias{hz21} the median quantities for each GES field rather than the individual measurements.

Additionally, we also find a linear relation ($r_p$=0.88) between EW and $\Av$ derived from the \citetalias{SFD} map with a calibration 
by \citet{SF11}, except for the GIBS field \#7 located at $(\ell,\,b)\,{=}\,(8.5^{\circ},\,{-}1.9^{\circ}),$ where the corresponding 
$\Av$ is clearly overestimated. This outlier is due to the fact that at low galactic latitudes, the \citetalias{SFD} map may 
not be reliable. Indeed, a comparison with \citetalias{Surot2020} reveals much lower extinction values for this field ($\Av\,{\sim}\,4\,$mag). 
The linear fit yields a coefficient of $\Av/{\rm EW}\,{=}\,10.733\,{\pm}\,0.972$. With the relation in NIR band $\EJKs/{\rm EW}\,{=}\,1.884$, 
we obtain $\EJKs/\Av\,{=}\,0.176$, slightly higher than the ratio of 0.170 predicted by the CCM model \citep{CCM89} with 
$\Rv\,{=}\,3.1$. On the other hand, the tight correlation with extinction makes DIB\,$\lambda$8620 a powerful tracer of any possible 
extinction law variation for different lines of sight. However, \citet{Krelowski18} argued the variation of the ${\rm EW}/\EBV$ 
ratio for DIB\,$\lambda$8620 by two stars HD 204827 and HD 219287 which have similar $\EBV$ but very different DIB profiles (see Fig. 
11 in his paper). 

\begin{figure}
    \centering
    \includegraphics[width=8cm]{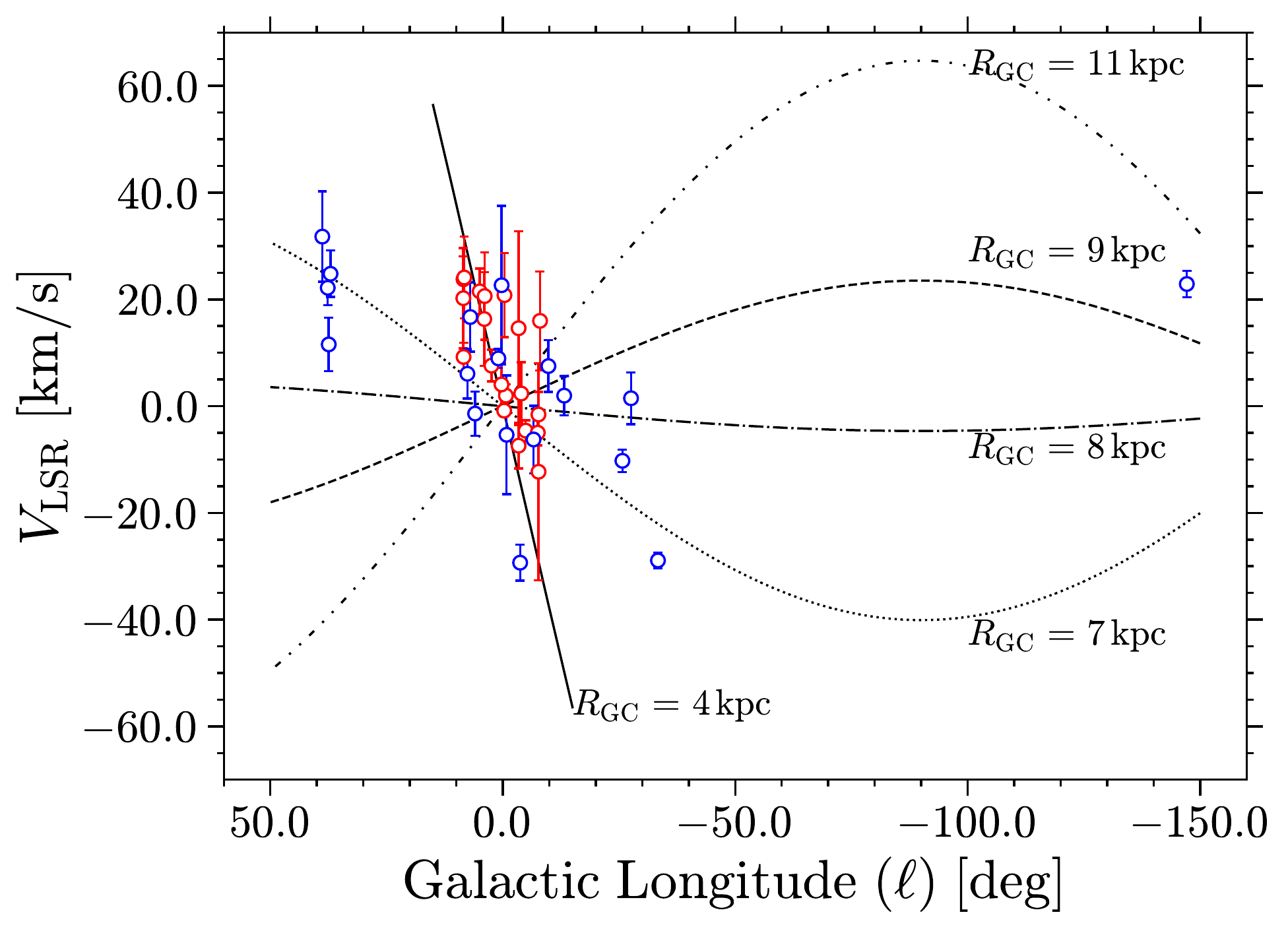}
    \caption{Longitude--velocity ($\ell-\Vlsr$) diagram for the GIBS (red) and GES (blue) fields, respectively.
    The circles indicate the median $\Vlsr$ and standard error of the mean for each field.
Velocity curves calculated by Model A5 in \citet{Reid19} for different galactocentric distances 
    ($\Rgc$) are overplotted.}
    \label{fig:VLSR}
\end{figure}

\begin{figure*}
    \centering
    \includegraphics[width=8.4cm]{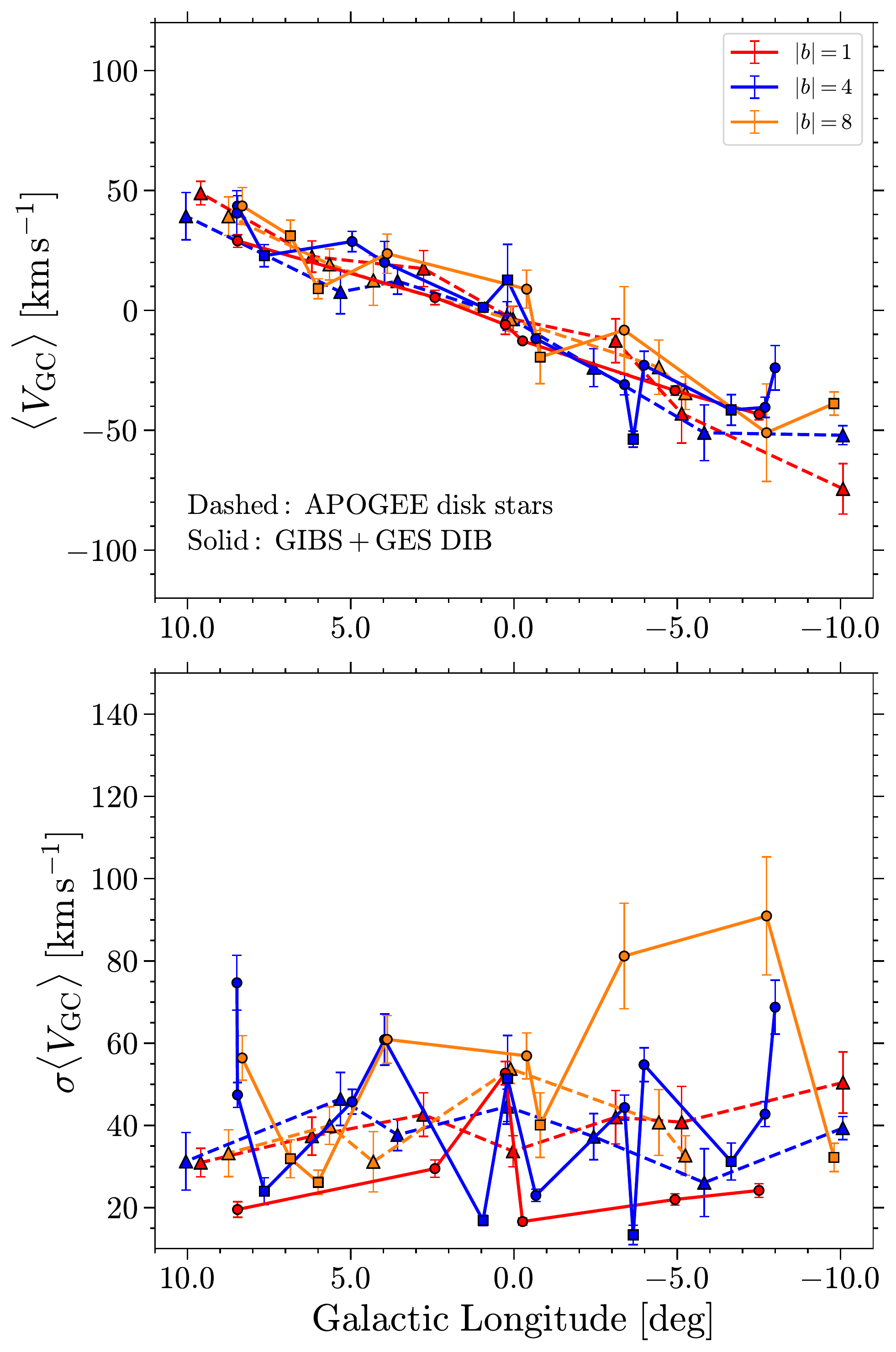}
    \includegraphics[width=8.4cm]{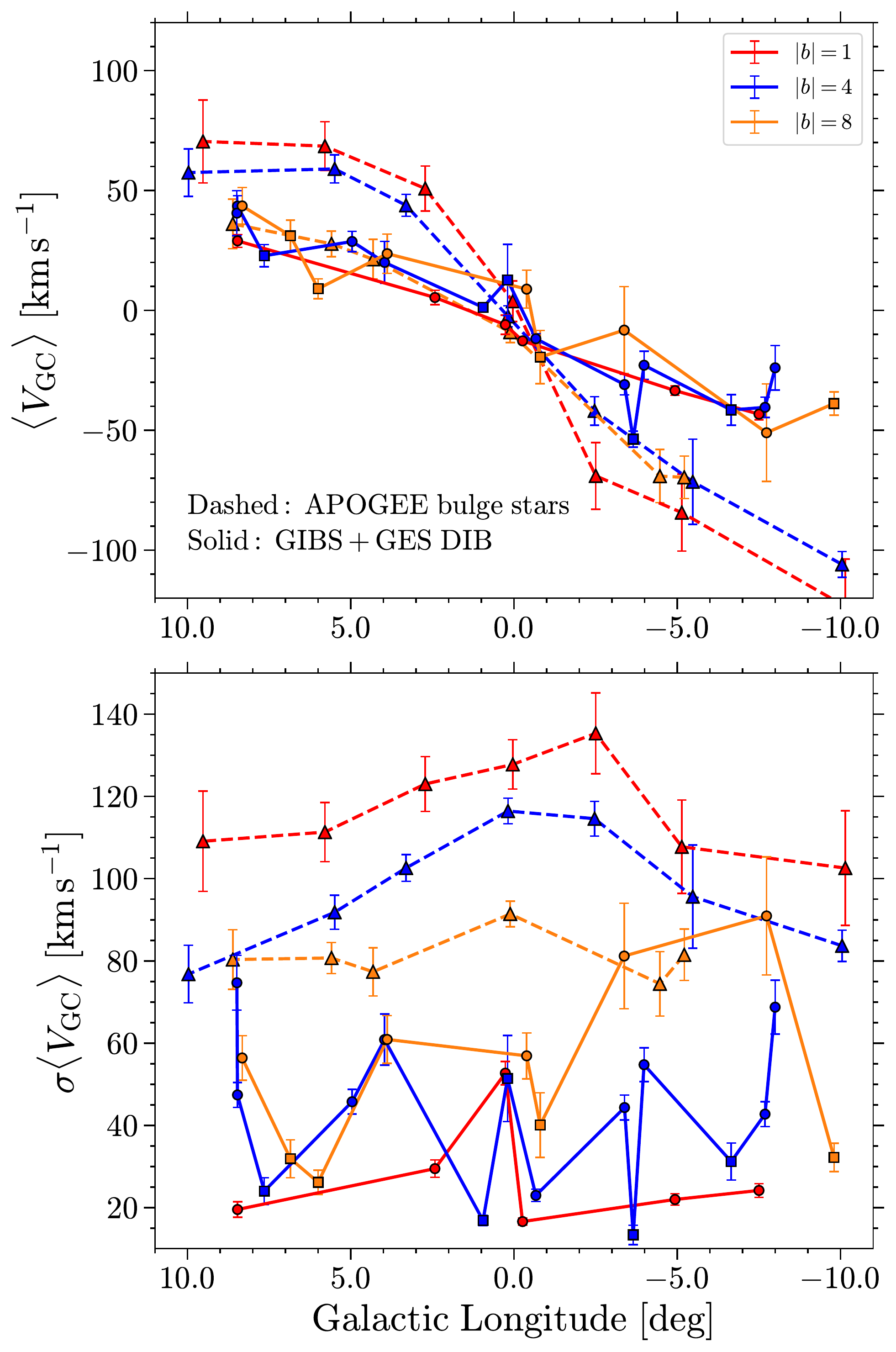}
    \caption{Median galactocentric radial velocity ({\it top}) and velocity dispersion ({\it bottom}) calculated in GIBS fields (circles), 
    GES fields (squares), and APOGEE bins (triangles), as a function of Galactic longitude, for different latitudes marked with 
    different colors. The APOGEE stars used in the {\it left panels} are in the disk with line-of-sight distances smaller than 3\,kpc.
    Bulge stars ($\Rgc\,{\leqslant}\,3$\,kpc) are used in {\it right panels}.}
    \label{fig:VGC}
\end{figure*}

\section{Kinematics of the DIB carrier} \label{sec:velocity}

\subsection{Rest-frame wavelength} \label{subsec:lambda0}

As the most important observational parameter, the precise DIB rest-frame wavelength ($\lambda_0$) is required for inferring the 
kinematic information of the DIB carrier. Although without a physical identification of the carrier, $\lambda_0$ can be determined 
with the empirical assumption that the radial velocity toward the Galactic center or the Galactic anti-center (see \citealt{Zasowski15}) 
is essentially null. Similarly to \citetalias{hz21}, 603 DIBs with $-3^{\circ}\,{<}\,b\,{<}\,3^{\circ}$ and $-6^{\circ}\,{<}\,\ell\,{<}\,3^{\circ}$ 
are selected from our pure GIBS sample which includes only the most reliable measurements (i.e., the error in EW is small). We obtained 
a median central wavelength in the heliocentric frame of $8620.52\,{\pm}\,0.36$\,{\AA}, close to the results of \citetalias{hz21} and 
\citet{Munari08}.  

However, both of the previous works did not consider the effect of the solar motion toward the Galactic center. Therefore, the rest-frame 
wavelength in the Local Standard of Rest (LSR) should be derived as $\lambda_0\,{=}\,\frac{c}{c-\Usun} \cdot C_{\rm obs,}$ 
where $c$ is the speed of light, $C_{\rm obs}$ the measured central wavelength in the heliocentric frame, and $\Usun$ the solar motion 
toward the Galactic center. In this work, we assume $\Usun\,{=}\,10.6\,{\pm}\,1.2\,\kms$ \citep[][Model A5]{Reid19} which gives a 
$\lambda_0\,{=}\,8620.83$\,{\AA} that is in good agreement with the value of 8620.79\,{\AA} in \citet{Galazutdinov00}. However, the 
solar motion $\Usun$ can vary between $10.1\,{\pm}\,1.0\,\kms$ \citep{Mroz19}, $10.3\,\kms$ \citep{Bovy12}, $10.6\,{\pm}\,1.2\,\kms$ 
\citep{Reid19}, $10.7\,{\pm}\,1.8\,\kms$ \citep{Reid14}, and $11.1\,\kms$ \citep{RB04,Schonrich10}. For an exhaustive summary 
of the measurements of the solar motion, we refer the reader to \citet{Wang2021}. We note that a difference of $\Delta \Usun\,{=}\,1\,\kms$ causes 
an error of $\sim$0.03\,{\AA} in $\lambda_0,$ while the typical error of $C_{\rm obs}$ in the fit is about 0.36\,{\AA}.

\subsection{Distribution of the carrier velocities} \label{subsec:VLSR}

We investigate here the radial velocity of the DIB carrier with respect to the LSR, $\Vlsr=\Vhc+\vec{V_{\sun}}\,{\cdot}\,\vec{A}$, 
where $\vec{V_{\sun}}=(10.6,10.7,7.6)\,\kms$ is the solar motion fitted by Model A5 of \citet{Reid19}, and 
$\vec{A}\,{=}\,(\cos(b)\cos(\ell),\cos(b)\sin(\ell),\sin(b))$ is the directional array of the DIB carrier. Figure \ref{fig:VLSR} presents 
$\Vlsr$ as a function of Galactic longitude for the GIBS (red circles) and GES (blue circles) fields, respectively, where we show 
the median $\Vlsr$ in each field. The error bars show the standard error of the mean (${\rm SEM}\,{=}\,\sigma/\sqrt{N}$, $N$ is the 
sample size) in each field. Indicated are Galactic rotation curves computed by Model A5 in \citet{Reid19} with different galactocentric 
radii ($\Rgc$).

Limited by the available sightlines, we cannot find a clear Galactic rotation from the GIBS$/$GES fields as seen in \citet[][Figure 8]{Zasowski15}.
Moreover, fields with $|\ell|\,{\leqslant}\,10^{\circ}$ present a large velocity dispersion caused by both the fitting errors (an 
error of 0.36\,{\AA} in $C_{\rm obs}$ amounts to $\Delta \Vlsr\,{\sim}\,10\,\kms$) and the velocity crowding \citep{Wenger18}. 
In Sect. \ref{sec:distance}, we apply two Galactic rotation models in order to derive kinematic distance for the DIB carrier.


\subsection{Galactocentric velocity and velocity dispersion} \label{subsec:VGC}

One of the known kinematic characteristics of the Galactic boxy/peanut bulge is its cylindrical rotation, which has already been
investigated in many studies \citep[see e.g.,][]{Ness13b,Zoccali14,Alvaro20}. This rotation curve is steeper in the bulge than for 
the Galactic disk \citep[see e.g.,][]{Howard09,Shen10,Zoccali14}. In contrast, the velocity dispersion in the bulge is higher with 
respect to the Galactic disk (by a factor of two or more), indicating more isotropic kinematics. Using these kinematic properties, we 
can determine if the DIB carriers are associated with the background stars or in the foreground disk. Here, we investigate the validity 
of this assumption. We use the APOGEE DR16 data set \citep{Majewski2017} as a comparison sample, and in particular the Galactic bulge 
sample from \citet{Alvaro20} with $|\ell|\,{\leqslant}\,11^{\circ}$, for three different Galactic latitude bins $|b|\,{=}\,1^{\circ},\,4^{\circ},\,8^{\circ}$. 
As shown by \citet{Alvaro20}, a simple cut at $\Rgc\,{\leqslant}\,3.5$\,kpc ensures a reliable bulge sample, which we applied.

Furthermore, we define a typical ``Galactic disk'' sample of APOGEE within the same longitude and latitude range as the bulge sample 
but with line-of-sight distances within 3\,kpc. The distances of our APOGEE stars were calculated by the same method used in \citet{Alvaro20} 
using the spectro-photometric distances together with PARSEC isochrones \citep{Alvaro17}. $\Vhc$ has been transformed to Galactocentric 
velocities ($\Vgc$) using the following formula \citep[e.g.,][]{Ness13b,Zoccali14}, where $(\ell,b)$ are the galactic coordinates:

\begin{equation} \label{eq:Vgc}
\begin{split}
\Vgc = \Vhc & + 220\,{\rm sin}(\ell)\,{\rm cos}(b) \\
            & + 16.5[{\rm sin}(b)\,{\rm sin(25)} + {\rm cos}(b)\,{\rm cos}(25)\,{\rm cos}(\ell-53)]
\end{split}
.\end{equation}

Figure \ref{fig:VGC} shows the rotation curves and velocity dispersion for the APOGEE sample (dashed lines) as well as our DIB 
measurements (solid lines). The APOGEE stars at different latitudes are further separated into eight equal longitude bins, where the 
median $\Vgc$ in each bin is used. Our full sample (GES and GIBS) has been divided into three groups with $|b|\,{<}\,3^{\circ}$, 
$3^{\circ}\,{<}\,|b|\,{<}\,7^{\circ}$ and $7^{\circ}\,{<}\,|b|\,{<}\,10^{\circ}$. The right panel of Fig. \ref{fig:VGC} shows the 
comparison between the bulge sample of APOGEE stars and our DIB samples. It is evident that in terms of rotation and velocity dispersion, 
our DIB measurements do not follow the general bulge characteristics of the stars. This is most striking in the velocity dispersion 
where the DIBs show more than a factor of two smaller velocity dispersion with respect to the bulge stars. In the left panel of Fig. 
\ref{fig:VGC}, we see, on the other hand, a much better agreement, with the disk sample from APOGEE showing similar velocity dispersion 
and rotation velocities. We therefore conclude that the DIB carriers located inside the Galactic disk ($4\,{<}\,\Rgc\,{<}\,11$\,kpc) 
could be far away from the background stars and much closer to us.

\section{Distance of the DIB carrier} \label{sec:distance}

As an interstellar feature, the DIB profile measured in the spectrum of a background star is the result of an integration of the 
DIB carrier between the observer and the star. The distance of the background star is then an upper limit on the typical distance 
of the DIB carrier along the sight line \citep{Zasowski15}. Based on hundreds of thousands spectra, \citet{Kos14} and \citet{Zasowski15} 
built 3D intensity maps for DIB\,$\lambda$8620 and DIB\,$\lambda$15273, respectively, using stellar distances and tracing the cumulation 
and variation of the DIB carriers along substantial sight lines. Their pioneering works encourage the following studies with 
large spectroscopic surveys, while in this work we attempt to estimate the distance of the DIB carrier more ``directly'' 
by the carrier radial velocity and Galactic rotation model; that is, the kinematic distance.

\subsection{Kinematic distance} \label{subsec:kine-dist}

\begin{figure}
    \centering
    \includegraphics[width=8cm]{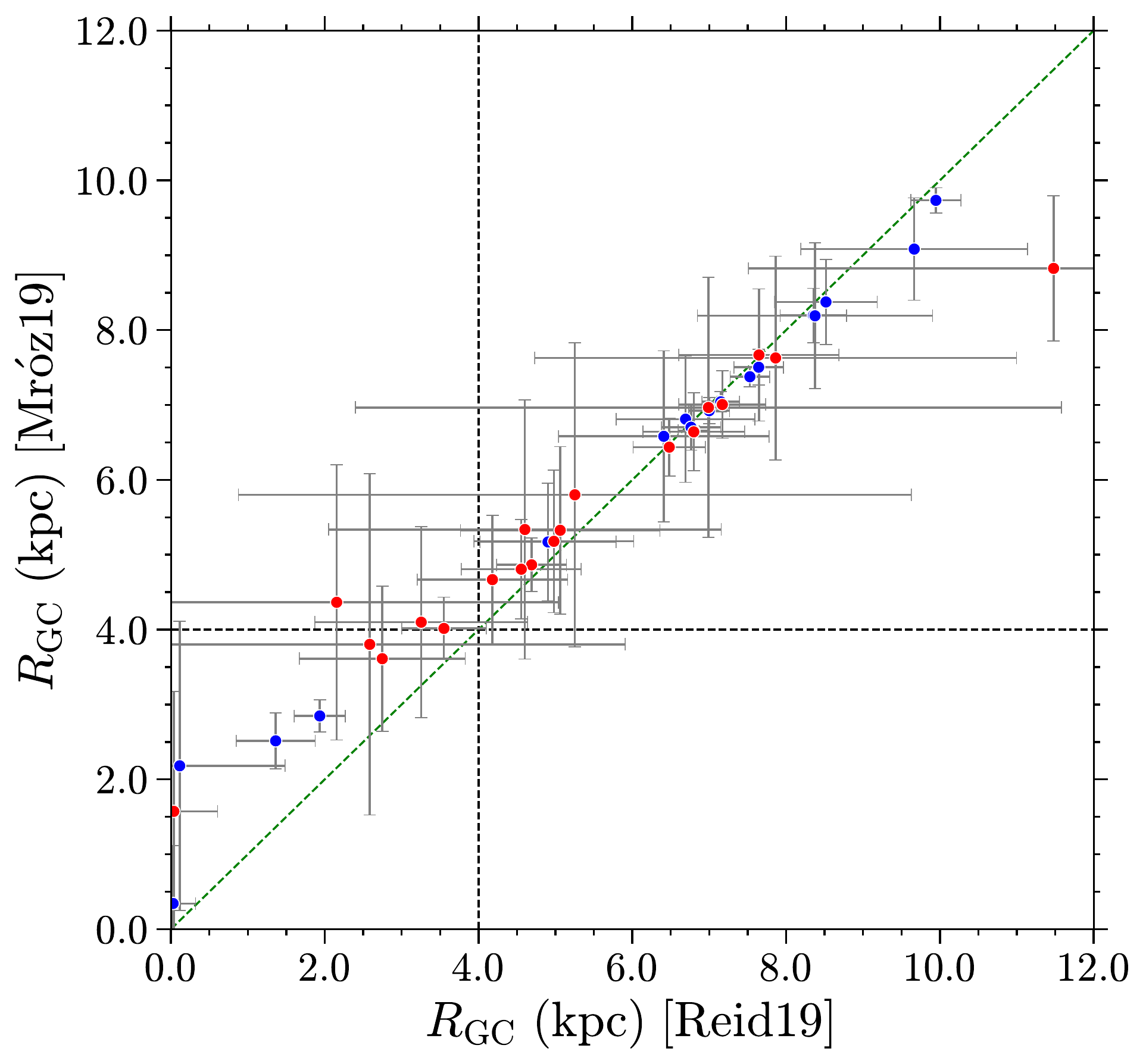}
    \caption{Comparison between kinematic distances derived by Model 2 in \citet{Mroz19} and Model A5 in \citet{Reid19}.
    The red and blue dots indicate the GIBS and GES fields, respectively. The dashed green line traces the one-to-one 
    correspondence. The dashed black lines indicate $\Rgc=4$\,kpc.}
    \label{fig:dib-Kdist}
\end{figure}

We calculated the kinematic distance of the DIB carrier using the median $\Vlsr$ in each field together with two different Galactic
rotation models. One is a two-parameter ``universal'' rotation model \citep[Model A5 in][]{Reid19}, the other is a linear rotation 
model \citep[Model 2 in][]{Mroz19}. For Model A5 in \citet{Reid19}, the kinematic distance is computed with a Monte Carlo method 
introduced by \citet{Wenger18}, using their python package {\it kd}\footnote{\url{https://github.com/tvwenger/kd}} and considering 
the error of $\Vlsr$ as well as the parameter uncertainty of the rotation model. For Model 2 in \cite{Mroz19}, only the error 
of $\Vlsr$ is considered. Figure \ref{fig:dib-Kdist} shows the comparison between the estimated galactocentric distances $\Rgc$ 
derived by the two models for our sample, which give consistent results with slightly larger distances for \citet{Mroz19} 
within $4\,{<}\,\Rgc\,{<}\,6$\,kpc, while these are smaller for $\Rgc\,{>}\,9$\,kpc. As both \citet{Reid19} and \citet{Mroz19} fit the models 
mainly within the 4--15\,kpc range, kinematic distances in the inner Galaxy with $\Rgc\,{<}\,4$\,kpc are not reliable and should 
not be used. In addition, in the inner Galaxy the orbits of the stars are highly eccentric (see e.g., \citealt{Alvaro20}), 
causing large errors for the kinematic distances.

We also calculated the line-of-sight distance:

\begin{equation} \label{eq:dlos}
    \dlos = \Rsun \cdot {\rm cos}(\ell) \pm \sqrt{\Rgc^2 - \Rsun^2 \cdot {\rm sin^2}(\ell)},
\end{equation} 

\noindent where $\Rsun=8.15$\,kpc \citep{Reid19}, $\ell$ is the median Galactic latitude of each field, and for $\Rgc$ we 
use the results from the model of \citet{Reid19}. The calculation was also completed by the {\it kd} package. For the inner Galaxy, 
Eq. \ref{eq:dlos} gives two possible solutions, of which we always chose the closest one. In case where the solution is negative or has no 
rational solution, the result has been dropped.

\begin{figure}
    \centering
    \includegraphics[width=8.4cm]{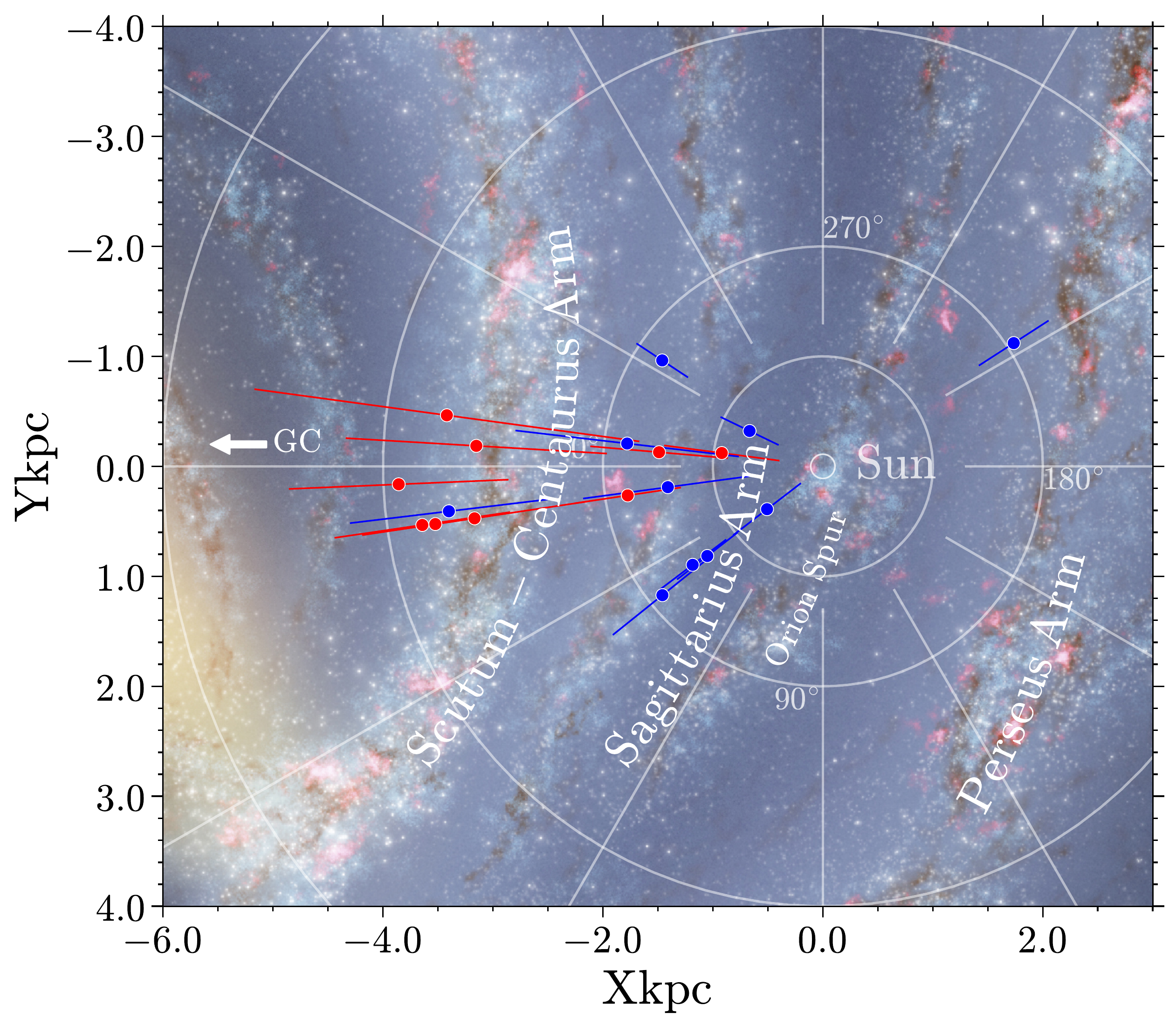}
    \caption{Face-on view of the spatial distribution of nine GIBS and ten GES fields with valid $\dlos$; i.e.,
    the kinematic distance of the DIB carrier calculated by the field-median radial velocity and model A5 in \citet{Reid19}, 
    overplotted with the Milky Way sketch created by Robert Hurt and Robert Benjamin \citep{Churchwell09}. The Galactic center 
    is located at $(-8,0)$. Red and blue dots indicate the GIBS and GES fields, respectively.}
    \label{fig:dib-KD-mw}
\end{figure}

Finally, we obtained $\Rgc\,{>}\,4$\,kpc for 14 GIBS and 14 GES fields. Nine GIBS and ten GES fields had valid $\dlos$. These 
measurements, together with their uncertainties, are listed in Table \ref{tab:field}. A face-on view of the distribution of the fields 
with valid $\dlos$ is shown in Fig. \ref{fig:dib-KD-mw}. The fields within $|\ell|\,{<}\,10^{\circ}$ experience large uncertainties 
than the fields outside. Seven fields are located within or beyond the Scutum--Centaurus Arm, while 11 other fields are around the Sagittarius 
Arm and the Orion Spur. The only field toward the Galactic anti-center at $(\ell,\,b)=({-}147.2^{\circ},\,{-}2.0^{\circ})$ almost reaches 
the edge of the Perseus Arm.

\subsection{Comparison with stellar distances} \label{subsec:star-dist}

The distances to the background stars are the upper limit on the carrier distances and can be used to test the reliability of our 
distance measurements. As the targets in the GIBS sample are RC stars, we can calculate their distance assuming $M_{\rm K_S}\,{=}\,-1.61$\,mag 
\citep{Ruiz-Dern18} and $(J-K_{\rm S})_0\,{=}\,0.674$\,mag \citep{Gonzalez12}. For the GES sample, we used spectro-photometric distances 
(see Sect.~\ref{sec:data}). For each field, we calculated the median distance together with its standard deviation. The results 
are shown in Table \ref{tab:field} for field medians and Table \ref{tab:ges-dib} for individual stars (some examples; full catalog 
can be accessed online). For a test, we also cross-matched the GES sample with the catalog of {\it Gaia}--EDR3 \citep{Gaia-EDR32020} 
within 1\arcsec and used the photogeometric distances estimated by \citet{Bailer-Jones2021} from the {\it Gaia} parallaxes. The
resulting median distances for each field are consistent with our calculations within the distance uncertainties.

\begin{figure}
    \centering
    \includegraphics[width=8cm]{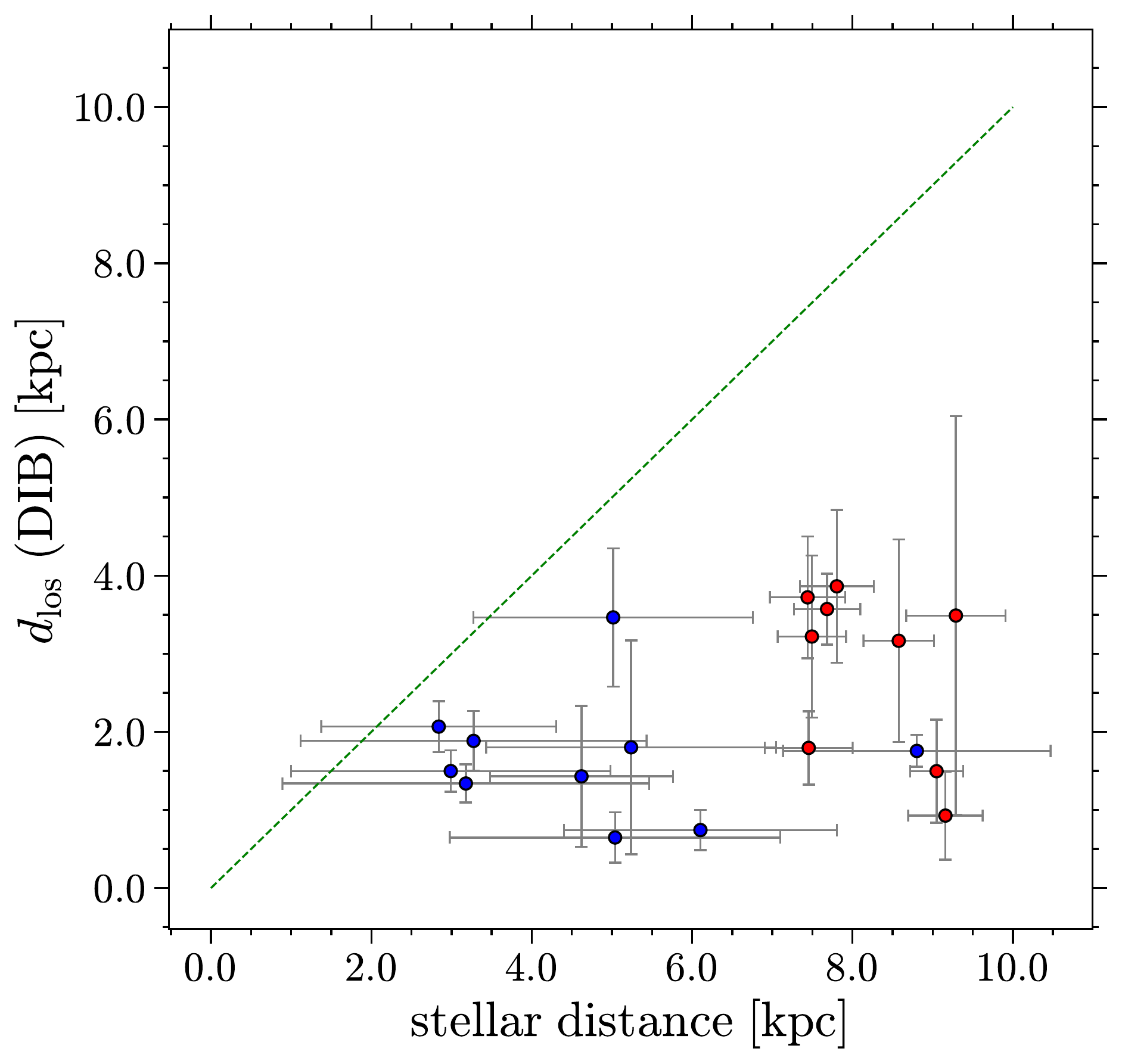}
    \caption{Comparison between stellar distance and the kinematic distance of the DIB carrier for nine GIBS (red dots) 
    and ten GES fields (blue dots). The dashed green line traces the one-to-one correspondence.}
    \label{fig:ds-compare}
\end{figure}

Figure \ref{fig:ds-compare} displays the comparison between the median stellar distances and the kinematic distance of the 
DIB carrier for nine GIBS and ten GES fields with valid $\dlos$. All of the points lie below the identity line (dashed green), 
indicating all the estimated carrier distances are smaller than the stellar distances. This is not very surprising for GIBS 
as the RC stars in GIBS fields are mainly distributed in the Galactic bulge, and we required the nearer solution for $\dlos$. It is more 
interesting for GES fields, of which the median distances are much closer to the Sun, meaning that the kinematic distance of the DIB carrier is still 
smaller than the stellar distance. This confirms the reliability of the derived kinematic distance to some extent. While we still 
need to emphasize that kinematic distances are of high uncertainty in the direction of the Galactic center and the Galactic anti-center 
\citep[see e.g.,][]{Balser15,Wenger18} and need to be carefully used.

The comparison in Fig. \ref{fig:ds-compare} also demonstrates that the DIB carrier can be located much closer to the observer than the 
background stars. So, when we make use of DIBs as a tool to trace the ISM environments and Galactic structure, such as local ISM medium 
\citep{PP2020}, Galactic arms \citep{PL2019}, and Galactic warp \citep{Istiqomah2020}, target stars at distant zones and/or high 
latitudes require more attention.

\section{Comparison with the NIR DIB\texorpdfstring{$\,\lambda$}{l}15273} \label{sec:apo-compare}

The correlation between different DIBs is one of the most important methods to study the relations of their carriers and to find the 
common carrier for a set of DIBs (see \citealt{Elyajouri17,Elyajouri18}, \citealt{Sonnentrucker2018}, \citealt{Galazutdinov2020}, and 
\citealt{Bondar2020} for some recent studies). The tightest correlation is between DIB\,$\lambda$6196 and DIB\,$\lambda$6614 \citep{McCall10}, 
although the conclusion of their common origin still encounters some problems \citep{Krelowski16}. \citet{Elyajouri17} reported tight 
correlations between the strong NIR DIB at $1.5273\,\um$ and the weak DIBs in its vicinity ($\lambda$15617, $\lambda$15653, and 
$\lambda$15673), as well as some strong optical DIBs, proposing DIB\,$\lambda$15273 as a good tracer of the interstellar environments.

\begin{figure}
    \centering
    \includegraphics[width=8cm]{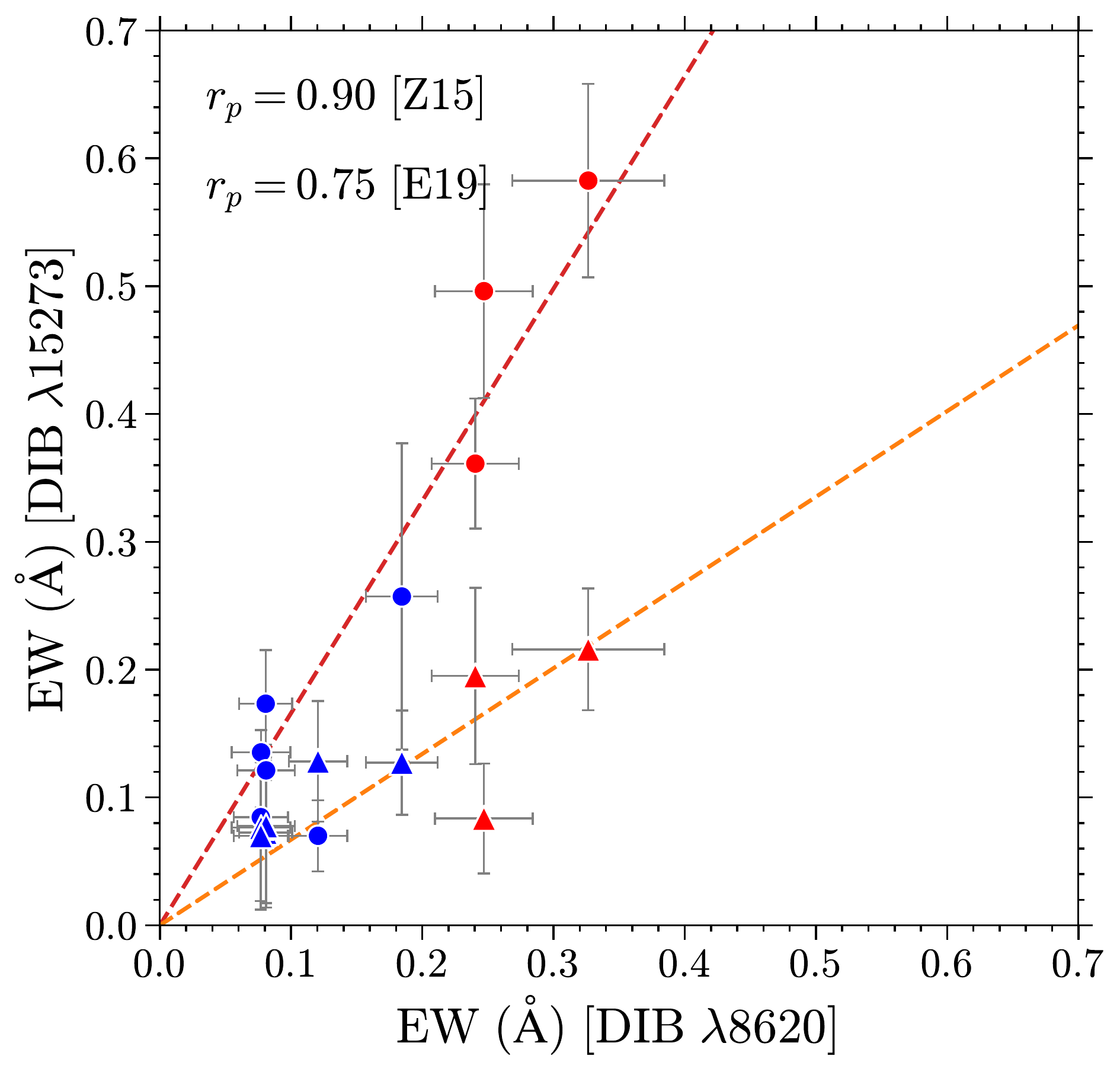}
    \caption{Median EW comparison between DIB\,$\lambda$8620 and DIB\,$\lambda$15273 for GIBS (red) and GES (blue) fields. Filled 
    circles indicate the DIB\,$\lambda$15273 measured by \citet{Zasowski15} and triangles by \citet{Elyajouri19}. The dashed red 
    line is the fit between the GIBS$/$GES sample and the \citet{Zasowski15} sample, and the dashed orange line between GIBS$/$GES 
    and \citet{Elyajouri19}. The error bars represent the standard deviation in each field. The Pearson correlation coefficients 
    ($r_p$) are also indicated.}
    \label{fig:apogee-dib}
\end{figure}

In this section, we make a simple comparison between DIB\,$\lambda$8620 measured in this work and DIB\,$\lambda$15273 in the APOGEE 
spectra measured by \citet{Zasowski15}. We accessed the full APOGEE DIB catalog with 49,474 
entries\footnote{\url{http://www.physics.utah.edu/~zasowski/APOGEE_DIB_Catalog.html}}, but cross-matched them with our GIBS$/$GES 
sample within 1{\arcsec}, where only 16 common targets were found (GES). This is due to the fact that APOGEE and GIBS$/$GES trace 
different stellar populations in their target selection; that is GIBS trace RC stars, while APOGEE traces brighter and cooler giants on the RGB. 
Therefore, we selected the APOGEE DIBs based on the same fields with respect to GIBS and GES and compared their median EW in each 
field. In total, we find six GES and three GIBS fields matching the APOGEE footprint. Figure \ref{fig:apogee-dib} shows the comparison 
between the EW of the two DIBs where a linear relation between the two carriers can be found ($r_p\,{=}\,0.90$). Clearly, more observations 
of these two DIB carriers spanning a larger EW range are needed in order to draw firmer conclusions. A linear fit to all 
the fields yields a ratio of $\rm EW\,\lambda 15273 / EW\,\lambda 8620\,{=}\,1.411\,{\pm}\,0.242$, demonstrating that the DIB\,$\lambda$15273 
is stronger than DIB\,$\lambda$8620. 

It should be noted that \citet{Elyajouri19} reported measurements\footnote{Data access: \url{https://cdsarc.unistra.fr/viz-bin/cat/J/A+A/628/A67}}
of DIB\,$\lambda$15273 that were systematically weaker than those in \citet{Zasowski15} due to the use of different stellar models. 
With their results, DIB\,$\lambda$8620 would be oppositely larger than the DIB\,$\lambda$15273, with a ratio of $\rm EW\,\lambda 15273 
/ EW\,\lambda 8620\,{=}\,0.670\,{\pm}\,0.132$ ($r_p\,{=}\,0.75$, see triangles in Fig. \ref{fig:apogee-dib}; the fields are selected with 
the same method for \citealt{Zasowski15}). However, this correlation was built only with the median values in very few common
fields. Thus, the present result is still a rough one, and further investigations with bigger data sets are expected. The correlation between 
DIB\,$\lambda$8620 and DIB\,$\lambda$15273, as well as other optical and infrared DIBs, will benefit from forthcoming new large 
spectroscopic data. The DIB\,$\lambda$8620 shows a promising diagnostic of the interstellar conditions and a tracer of the Galactic 
structure.

We derived a linear correlation between EW and $\Av$ with ${\rm EW}\,\lambda 8620/\Av\,{=}\,0.093$\,{\AA}\,$\rm mag^{-1}$ for our 
GIBS$/$GES fields (Sect. \ref{subsec:linear}). Compared to ${\rm EW}\,\lambda 15273/\Av\,{=}\,0.102$\,{\AA}\,$\rm mag^{-1}$ from 
\citet{Zasowski15}, we can obtain a ratio of 1.097 for $\rm EW\,\lambda 15273 / EW\,\lambda 8620$, which is about 22\% lower than the 
value from the direct EW comparison with \citet{Zasowski15}. This difference could be caused by the different extinction sources 
used in \citet{Zasowski15} and this work; \citet{Zasowski15} applied the RJCE method \citep{Majewski2011} for their extinction values, 
while we used the $\Av$ from the \citetalias{SFD} map with a calibration of \citet{SF11}.   

\citet{Zasowski15} roughly estimated the carrier abundance relative to hydrogen for DIB\,$\lambda$15273 as $N_{\rm DIB} / N_{\rm H} 
\sim 2.3 \times 10^{-11}/f$ with its relationship to $\Av$, $\lambda_0 = 15272.42$\,{\AA}, and a mean hydrogen-to-extinction relation 
of $N_{\rm H}/\Av = 2 \times 10^{21}\,{\rm cm^{-2}\, mag^{-1}}$ \citep{DL90}, where $f$ is the transition oscillator strength 
\citep[e.g.,][]{Spitzer78}. With our derived relation ${\rm EW}\,\lambda 8620/\Av = 0.093$\,{\AA}\,$\rm mag^{-1}$ and $\lambda_0 = 
8620.83$\,{\AA}, the carrier abundance for DIB\,$\lambda$8620 is estimated as $1.4 \times 10^{-11}/f$, slightly lower than the value 
of the DIB\,$\lambda$15273.

\section{Conclusions and summary} \label{sec:conclusion}

In this work, we successfully detected the DIB\,$\lambda$8620 in 760 GES spectra with $|b|\,{\leqslant}\,10^{\circ}$ and $\rm S/N\,{>}\,50$. 
Their EW, as well as depth, central wavelength, and width, were measured with a Gaussian profile. Our EWs were slightly smaller than 
those measured in \citet{Puspitarini15}, with a mean difference of 0.031\,{\AA} for 43 common targets. The linear relation between 
EW and $\EJKs$ ($r_p=0.91$) derived from field median values is highly consistent with the recommended correlation derived 
in \citetalias{hz21}.

Combined with a pure GIBS sample from \citetalias{hz21}, we confirmed a linear relation between EW and $\Av$ (\citetalias{SFD}) with 
$r_p=0.88$, using 2540 DIBs distributed in 38 fields. We obtained $\EJKs/\Av\,{=}\,0.176$ from their linear fit with EW, which was 
slightly higher than the ratio of 0.170 predicted by the CCM model \citep{CCM89}. Furthermore, the rest-frame wavelength of 
DIB\,$\lambda$8620 was redetermined as $\lambda_0=8620.83 \pm 0.36$\,{\AA} after the consideration of the solar motion.

We also studied the kinematics of the DIB carriers based on the median radial velocities in each field. Most of our fields distributed 
close to the Galactic center ($|\ell|\,{\leqslant}\,10^{\circ}$), thus they were crowded in the $\ell-\Vlsr$ diagram with large scatters. 
The $\ell-\Vgc$ diagram showed that the DIB carriers mainly occupied in the local Galactic disk as traced by a sample of APOGEE stars. 

Applying the Galactic rotation models \citep{Reid19,Mroz19}, we calculated the kinematic distances of the DIB carriers for 
each field and got valid line-of-sight distances ($\dlos$) for nine GIBS and ten GES fields. All derived $\dlos$ are smaller than 
the median distances to background stars in each field. It demonstrates that the DIB carriers can be located much closer to us than the 
background stars. Therefore, when we make use of target stars to build the integrated DIB map, we have to be careful with the distant- 
and/or high-latitude zones, as well as the region where stars are not well sampled.

For the first time, we roughly investigated the mutual correlation between DIB\,$\lambda$8620 measured in this work and DIB\,$\lambda$15273 
in \citet{Zasowski15} and \citet{Elyajouri19}, respectively, with three GIBS fields and six GES fields, resulting in a liner coefficient 
of $\rm EW\,\lambda 15273 / EW\,\lambda 8620\,{=}\,1.411\,{\pm}\,0.242$ for the measurements from \citet{Zasowski15} and $0.670\,{\pm}\,0.132$ 
for \citet{Elyajouri19}. The Pearson correlation coefficients are 0.90 and 0.75, respectively. The difference was caused by the use 
of different stellar templates for APOGEE spectra. The linear correlation suggested that DIB\,$\lambda$8620 may also correlate with 
other optical and infrared DIBs such as DIB\,$\lambda$15273, which can be used to trace the interstellar environments and Galactic structure 
with the measurements in large spectroscopic surveys.

\begin{acknowledgements}
  HZ is funded by the China Scholarship Council (No. 201806040200).
  MS acknowledges the Programme National de Cosmologie et Galaxies (PNCG) of CNRS/INSU, France, for financial support.
  ARA acknowledges support from FONDECYT through grant 3180203.
  The synthetic spectra grid calculations for GES targets have been performed with the high-performance computing facility SIGAMM, hosted by OCA.
\end{acknowledgements}

\bibliographystyle{aa}
\bibliography{references}

\begin{thebibliography}{108}
\expandafter\ifx\csname natexlab\endcsname\relax\def\natexlab#1{#1}\fi

\bibitem[{{Bailer-Jones} {et~al.}(2021){Bailer-Jones}, {Rybizki}, {Fouesneau},
  {Demleitner}, \& {Andrae}}]{Bailer-Jones2021}
{Bailer-Jones}, C.~A.~L., {Rybizki}, J., {Fouesneau}, M., {Demleitner}, M., \&
  {Andrae}, R. 2021,
  \href{http://dx.doi.org/10.3847/1538-3881/abd806}{\color{magenta}\aj},
  \href{https://ui.adsabs.harvard.edu/abs/2021AJ....161..147B}{161, 147}

\bibitem[{{Balser} {et~al.}(2015){Balser}, {Wenger}, {Anderson}, \&
  {Bania}}]{Balser15}
{Balser}, D.~S., {Wenger}, T.~V., {Anderson}, L.~D., \& {Bania}, T.~M. 2015,
  \href{http://dx.doi.org/10.1088/0004-637X/806/2/199}{\color{magenta}\apj},
  \href{https://ui.adsabs.harvard.edu/abs/2015ApJ...806..199B}{806, 199}

\bibitem[{{Bergemann}(2021, in prep)}]{Bergemann2021pre}
{Bergemann}. 2021, in prep

\bibitem[{{Bijaoui}(2012)}]{Bijaoui2012}
{Bijaoui}, A. 2012, in Seventh Conference on Astronomical Data Analysis, ed.
  J.-L. {Starck} \& C.~{Surace},
  \href{https://ui.adsabs.harvard.edu/abs/2012ada..confE...2B}{2}

\bibitem[{{Bondar}(2020)}]{Bondar2020}
{Bondar}, A. 2020,
  \href{http://dx.doi.org/10.1093/mnras/staa1610}{\color{magenta}\mnras},
  \href{https://ui.adsabs.harvard.edu/abs/2020MNRAS.496.2231B}{496, 2231}

\bibitem[{{Bovy} {et~al.}(2012){Bovy}, {Allende Prieto}, {Beers}, {Bizyaev},
  {da Costa}, {Cunha}, {Ebelke}, {Eisenstein}, {Frinchaboy}, {Garc{\'\i}a
  P{\'e}rez}, {Girardi}, {Hearty}, {Hogg}, {Holtzman}, {Maia}, {Majewski},
  {Malanushenko}, {Malanushenko}, {M{\'e}sz{\'a}ros}, {Nidever}, {O'Connell},
  {O'Donnell}, {Oravetz}, {Pan}, {Rocha-Pinto}, {Schiavon}, {Schneider},
  {Schultheis}, {Skrutskie}, {Smith}, {Weinberg}, {Wilson}, \&
  {Zasowski}}]{Bovy12}
{Bovy}, J., {Allende Prieto}, C., {Beers}, T.~C., {et~al.} 2012,
  \href{http://dx.doi.org/10.1088/0004-637X/759/2/131}{\color{magenta}\apj},
  \href{https://ui.adsabs.harvard.edu/abs/2012ApJ...759..131B}{759, 131}

\bibitem[{{Campbell} {et~al.}(2015){Campbell}, {Holz}, {Gerlich}, \&
  {Maier}}]{Campbell15}
{Campbell}, E.~K., {Holz}, M., {Gerlich}, D., \& {Maier}, J.~P. 2015,
  \href{http://dx.doi.org/10.1038/nature14566}{\color{magenta}\nat},
  \href{https://ui.adsabs.harvard.edu/abs/2015Natur.523..322C}{523, 322}

\bibitem[{{Campbell} {et~al.}(2016{\natexlab{a}}){Campbell}, {Holz}, \&
  {Maier}}]{Campbell2016b}
{Campbell}, E.~K., {Holz}, M., \& {Maier}, J.~P. 2016{\natexlab{a}},
  \href{http://dx.doi.org/10.3847/2041-8205/826/1/L4}{\color{magenta}\apjl},
  \href{https://ui.adsabs.harvard.edu/abs/2016ApJ...826L...4C}{826, L4}

\bibitem[{{Campbell} {et~al.}(2016{\natexlab{b}}){Campbell}, {Holz}, {Maier},
  {Gerlich}, {Walker}, \& {Bohlender}}]{Campbell2016a}
{Campbell}, E.~K., {Holz}, M., {Maier}, J.~P., {et~al.} 2016{\natexlab{b}},
  \href{http://dx.doi.org/10.3847/0004-637X/822/1/17}{\color{magenta}\apj},
  \href{https://ui.adsabs.harvard.edu/abs/2016ApJ...822...17C}{822, 17}

\bibitem[{{Campbell} \& {Maier}(2018)}]{CM2018}
{Campbell}, E.~K. \& {Maier}, J.~P. 2018,
  \href{http://dx.doi.org/10.3847/1538-4357/aab963}{\color{magenta}\apj},
  \href{https://ui.adsabs.harvard.edu/abs/2018ApJ...858...36C}{858, 36}

\bibitem[{{Cardelli} {et~al.}(1989){Cardelli}, {Clayton}, \& {Mathis}}]{CCM89}
{Cardelli}, J.~A., {Clayton}, G.~C., \& {Mathis}, J.~S. 1989,
  \href{http://dx.doi.org/10.1086/167900}{\color{magenta}\apj},
  \href{https://ui.adsabs.harvard.edu/abs/1989ApJ...345..245C}{345, 245}

\bibitem[{{Chen} {et~al.}(2013){Chen}, {Lallement}, {Babusiaux}, {Puspitarini},
  {Bonifacio}, \& {Hill}}]{ChenHC13}
{Chen}, H.~C., {Lallement}, R., {Babusiaux}, C., {et~al.} 2013,
  \href{http://dx.doi.org/10.1051/0004-6361/201220413}{\color{magenta}\aap},
  \href{https://ui.adsabs.harvard.edu/abs/2013A&A...550A..62C}{550, A62}

\bibitem[{{Churchwell} {et~al.}(2009){Churchwell}, {Babler}, {Meade},
  {Whitney}, {Benjamin}, {Indebetouw}, {Cyganowski}, {Robitaille}, {Povich},
  {Watson}, \& {Bracker}}]{Churchwell09}
{Churchwell}, E., {Babler}, B.~L., {Meade}, M.~R., {et~al.} 2009,
  \href{http://dx.doi.org/10.1086/597811}{\color{magenta}\pasp},
  \href{https://ui.adsabs.harvard.edu/abs/2009PASP..121..213C}{121, 213}

\bibitem[{{Cordiner} {et~al.}(2017){Cordiner}, {Cox}, {Lallement}, {Najarro},
  {Cami}, {Gull}, {Foing}, {Linnartz}, {Lindler}, {Proffitt}, {Sarre}, \&
  {Charnley}}]{Cordiner17}
{Cordiner}, M.~A., {Cox}, N.~L.~J., {Lallement}, R., {et~al.} 2017,
  \href{http://dx.doi.org/10.3847/2041-8213/aa78f7}{\color{magenta}\apjl},
  \href{https://ui.adsabs.harvard.edu/abs/2017ApJ...843L...2C}{843, L2}

\bibitem[{{Cordiner} {et~al.}(2019){Cordiner}, {Linnartz}, {Cox}, {Cami},
  {Najarro}, {Proffitt}, {Lallement}, {Ehrenfreund}, {Foing}, {Gull}, {Sarre},
  \& {Charnley}}]{Cordiner19}
{Cordiner}, M.~A., {Linnartz}, H., {Cox}, N.~L.~J., {et~al.} 2019,
  \href{http://dx.doi.org/10.3847/2041-8213/ab14e5}{\color{magenta}\apjl},
  \href{https://ui.adsabs.harvard.edu/abs/2019ApJ...875L..28C}{875, L28}

\bibitem[{{Cox} {et~al.}(2007){Cox}, {Boudin}, {Foing}, {Schnerr}, {Kaper},
  {Neiner}, {Henrichs}, {Donati}, \& {Ehrenfreund}}]{Cox07}
{Cox}, N.~L.~J., {Boudin}, N., {Foing}, B.~H., {et~al.} 2007,
  \href{http://dx.doi.org/10.1051/0004-6361:20065278}{\color{magenta}\aap},
  \href{https://ui.adsabs.harvard.edu/abs/2007A&A...465..899C}{465, 899}

\bibitem[{{Cox} {et~al.}(2014){Cox}, {Cami}, {Kaper}, {Ehrenfreund}, {Foing},
  {Ochsendorf}, {van Hooff}, \& {Salama}}]{Cox14}
{Cox}, N.~L.~J., {Cami}, J., {Kaper}, L., {et~al.} 2014,
  \href{http://dx.doi.org/10.1051/0004-6361/201323061}{\color{magenta}\aap},
  \href{https://ui.adsabs.harvard.edu/abs/2014A&A...569A.117C}{569, A117}

\bibitem[{{Cox} {et~al.}(2011){Cox}, {Ehrenfreund}, {Foing}, {D'Hendecourt},
  {Salama}, \& {Sarre}}]{Cox11}
{Cox}, N.~L.~J., {Ehrenfreund}, P., {Foing}, B.~H., {et~al.} 2011,
  \href{http://dx.doi.org/10.1051/0004-6361/201016365}{\color{magenta}\aap},
  \href{https://ui.adsabs.harvard.edu/abs/2011A&A...531A..25C}{531, A25}

\bibitem[{{Damineli} {et~al.}(2016){Damineli}, {Almeida}, {Blum}, {Damineli},
  {Navarete}, {Rubinho}, \& {Teodoro}}]{Damineli16}
{Damineli}, A., {Almeida}, L.~A., {Blum}, R.~D., {et~al.} 2016,
  \href{http://dx.doi.org/10.1093/mnras/stw2122}{\color{magenta}\mnras},
  \href{https://ui.adsabs.harvard.edu/abs/2016MNRAS.463.2653D}{463, 2653}

\bibitem[{{de Laverny} {et~al.}(2013){de Laverny}, {Recio-Blanco}, {Worley},
  {De Pascale}, {Hill}, \& {Bijaoui}}]{de-Laverny2013}
{de Laverny}, P., {Recio-Blanco}, A., {Worley}, C.~C., {et~al.} 2013, The
  Messenger, \href{https://ui.adsabs.harvard.edu/abs/2013Msngr.153...18D}{153,
  18}

\bibitem[{{de Laverny} {et~al.}(2012){de Laverny}, {Recio-Blanco}, {Worley}, \&
  {Plez}}]{de-Laverny2012}
{de Laverny}, P., {Recio-Blanco}, A., {Worley}, C.~C., \& {Plez}, B. 2012,
  \href{http://dx.doi.org/10.1051/0004-6361/201219330}{\color{magenta}\aap},
  \href{https://ui.adsabs.harvard.edu/abs/2012A&A...544A.126D}{544, A126}

\bibitem[{{Dickey} \& {Lockman}(1990)}]{DL90}
{Dickey}, J.~M. \& {Lockman}, F.~J. 1990,
  \href{http://dx.doi.org/10.1146/annurev.aa.28.090190.001243}{\color{magenta}\araa},
  \href{https://ui.adsabs.harvard.edu/abs/1990ARA&A..28..215D}{28, 215}

\bibitem[{{Eisenstein} {et~al.}(2011){Eisenstein}, {Weinberg}, {Agol},
  {Aihara}, {Allende Prieto}, {Anderson}, {Arns}, {Aubourg}, {Bailey},
  {Balbinot}, {Barkhouser}, {Beers}, {Berlind}, {Bickerton}, {Bizyaev},
  {Blanton}, {Bochanski}, {Bolton}, {Bosman}, {Bovy}, {Brandt}, {Breslauer},
  {Brewington}, {Brinkmann}, {Brown}, {Brownstein}, {Burger}, {Busca},
  {Campbell}, {Cargile}, {Carithers}, {Carlberg}, {Carr}, {Chang}, {Chen},
  {Chiappini}, {Comparat}, {Connolly}, {Cortes}, {Croft}, {Cunha}, {da Costa},
  {Davenport}, {Dawson}, {De Lee}, {Porto de Mello}, {de Simoni}, {Dean},
  {Dhital}, {Ealet}, {Ebelke}, {Edmondson}, {Eiting}, {Escoffier}, {Esposito},
  {Evans}, {Fan}, {Femen{\'\i}a Castell{\'a}}, {Dutra Ferreira}, {Fitzgerald},
  {Fleming}, {Font-Ribera}, {Ford}, {Frinchaboy}, {Garc{\'\i}a P{\'e}rez},
  {Gaudi}, {Ge}, {Ghezzi}, {Gillespie}, {Gilmore}, {Girardi}, {Gott}, {Gould},
  {Grebel}, {Gunn}, {Hamilton}, {Harding}, {Harris}, {Hawley}, {Hearty},
  {Hennawi}, {Gonz{\'a}lez Hern{\'a}ndez}, {Ho}, {Hogg}, {Holtzman},
  {Honscheid}, {Inada}, {Ivans}, {Jiang}, {Jiang}, {Johnson}, {Jordan},
  {Jordan}, {Kauffmann}, {Kazin}, {Kirkby}, {Klaene}, {Knapp}, {Kneib},
  {Kochanek}, {Koesterke}, {Kollmeier}, {Kron}, {Lampeitl}, {Lang}, {Lawler},
  {Le Goff}, {Lee}, {Lee}, {Leisenring}, {Lin}, {Liu}, {Long}, {Loomis},
  {Lucatello}, {Lundgren}, {Lupton}, {Ma}, {Ma}, {MacDonald}, {Mack},
  {Mahadevan}, {Maia}, {Majewski}, {Makler}, {Malanushenko}, {Malanushenko},
  {Mand elbaum}, {Maraston}, {Margala}, {Maseman}, {Masters}, {McBride},
  {McDonald}, {McGreer}, {McMahon}, {Mena Requejo}, {M{\'e}nard},
  {Miralda-Escud{\'e}}, {Morrison}, {Mullally}, {Muna}, {Murayama}, {Myers},
  {Naugle}, {Neto}, {Nguyen}, {Nichol}, {Nidever}, {O'Connell}, {Ogando},
  {Olmstead}, {Oravetz}, {Padmanabhan}, {Paegert}, {Palanque-Delabrouille},
  {Pan}, {Pandey}, {Parejko}, {P{\^a}ris}, {Pellegrini}, {Pepper}, {Percival},
  {Petitjean}, {Pfaffenberger}, {Pforr}, {Phleps}, {Pichon}, {Pieri}, {Prada},
  {Price-Whelan}, {Raddick}, {Ramos}, {Reid}, {Reyle}, {Rich}, {Richards},
  {Rieke}, {Rieke}, {Rix}, {Robin}, {Rocha-Pinto}, {Rockosi}, {Roe},
  {Rollinde}, {Ross}, {Ross}, {Rossetto}, {S{\'a}nchez}, {Santiago}, {Sayres},
  {Schiavon}, {Schlegel}, {Schlesinger}, {Schmidt}, {Schneider}, {Sellgren},
  {Shelden}, {Sheldon}, {Shetrone}, {Shu}, {Silverman}, {Simmerer}, {Simmons},
  {Sivarani}, {Skrutskie}, {Slosar}, {Smee}, {Smith}, {Snedden}, {Stassun},
  {Steele}, {Steinmetz}, {Stockett}, {Stollberg}, {Strauss}, {Szalay},
  {Tanaka}, {Thakar}, {Thomas}, {Tinker}, {Tofflemire}, {Tojeiro}, {Tremonti},
  {Vargas Maga{\~n}a}, {Verde}, {Vogt}, {Wake}, {Wan}, {Wang}, {Weaver},
  {White}, {White}, {Wilson}, {Wisniewski}, {Wood-Vasey}, {Yanny}, {Yasuda},
  {Y{\`e}che}, {York}, {Young}, {Zasowski}, {Zehavi}, \& {Zhao}}]{Eisenstein11}
{Eisenstein}, D.~J., {Weinberg}, D.~H., {Agol}, E., {et~al.} 2011,
  \href{http://dx.doi.org/10.1088/0004-6256/142/3/72}{\color{magenta}\aj},
  \href{https://ui.adsabs.harvard.edu/abs/2011AJ....142...72E}{142, 72}

\bibitem[{{Elyajouri} \& {Lallement}(2019)}]{Elyajouri19}
{Elyajouri}, M. \& {Lallement}, R. 2019,
  \href{http://dx.doi.org/10.1051/0004-6361/201834452}{\color{magenta}\aap},
  \href{https://ui.adsabs.harvard.edu/abs/2019A&A...628A..67E}{628, A67}

\bibitem[{{Elyajouri} {et~al.}(2018){Elyajouri}, {Lallement}, {Cox}, {Cami},
  {Cordiner}, {Smoker}, {Farhang}, {Sarre}, \& {Linnartz}}]{Elyajouri18}
{Elyajouri}, M., {Lallement}, R., {Cox}, N.~L.~J., {et~al.} 2018,
  \href{http://dx.doi.org/10.1051/0004-6361/201833105}{\color{magenta}\aap},
  \href{https://ui.adsabs.harvard.edu/abs/2018A&A...616A.143E}{616, A143}

\bibitem[{{Elyajouri} {et~al.}(2017){Elyajouri}, {Lallement}, {Monreal-Ibero},
  {Capitanio}, \& {Cox}}]{Elyajouri17}
{Elyajouri}, M., {Lallement}, R., {Monreal-Ibero}, A., {Capitanio}, L., \&
  {Cox}, N.~L.~J. 2017,
  \href{http://dx.doi.org/10.1051/0004-6361/201630088}{\color{magenta}\aap},
  \href{https://ui.adsabs.harvard.edu/abs/2017A&A...600A.129E}{600, A129}

\bibitem[{{Fan} {et~al.}(2019){Fan}, {Hobbs}, {Dahlstrom}, {Welty}, {York},
  {Rachford}, {Snow}, {Sonnentrucker}, {Baskes}, \& {Zhao}}]{Fan19}
{Fan}, H., {Hobbs}, L.~M., {Dahlstrom}, J.~A., {et~al.} 2019,
  \href{http://dx.doi.org/10.3847/1538-4357/ab1b74}{\color{magenta}\apj},
  \href{https://ui.adsabs.harvard.edu/abs/2019ApJ...878..151F}{878, 151}

\bibitem[{{Friedman} {et~al.}(2011){Friedman}, {York}, {McCall}, {Dahlstrom},
  {Sonnentrucker}, {Welty}, {Drosback}, {Hobbs}, {Rachford}, \&
  {Snow}}]{Friedman2011}
{Friedman}, S.~D., {York}, D.~G., {McCall}, B.~J., {et~al.} 2011,
  \href{http://dx.doi.org/10.1088/0004-637X/727/1/33}{\color{magenta}\apj},
  \href{https://ui.adsabs.harvard.edu/abs/2011ApJ...727...33F}{727, 33}

\bibitem[{{Gaia Collaboration} {et~al.}(2021){Gaia Collaboration}, {Brown},
  {Vallenari}, {Prusti}, {de Bruijne}, {Babusiaux}, {Biermann}, {Creevey},
  {Evans}, {Eyer}, {Hutton}, {Jansen}, {Jordi}, {Klioner}, {Lammers},
  {Lindegren}, {Luri}, {Mignard}, {Panem}, {Pourbaix}, {Randich}, {Sartoretti},
  {Soubiran}, {Walton}, {Arenou}, {Bailer-Jones}, {Bastian}, {Cropper},
  {Drimmel}, {Katz}, {Lattanzi}, {van Leeuwen}, {Bakker}, {Cacciari},
  {Casta{\~n}eda}, {De Angeli}, {Ducourant}, {Fabricius}, {Fouesneau},
  {Fr{\'e}mat}, {Guerra}, {Guerrier}, {Guiraud}, {Jean-Antoine Piccolo},
  {Masana}, {Messineo}, {Mowlavi}, {Nicolas}, {Nienartowicz}, {Pailler},
  {Panuzzo}, {Riclet}, {Roux}, {Seabroke}, {Sordo}, {Tanga}, {Th{\'e}venin},
  {Gracia-Abril}, {Portell}, {Teyssier}, {Altmann}, {Andrae}, {Bellas-Velidis},
  {Benson}, {Berthier}, {Blomme}, {Brugaletta}, {Burgess}, {Busso}, {Carry},
  {Cellino}, {Cheek}, {Clementini}, {Damerdji}, {Davidson}, {Delchambre},
  {Dell'Oro}, {Fern{\'a}ndez-Hern{\'a}ndez}, {Galluccio}, {Garc{\'\i}a-Lario},
  {Garcia-Reinaldos}, {Gonz{\'a}lez-N{\'u}{\~n}ez}, {Gosset}, {Haigron},
  {Halbwachs}, {Hambly}, {Harrison}, {Hatzidimitriou}, {Heiter},
  {Hern{\'a}ndez}, {Hestroffer}, {Hodgkin}, {Holl}, {Jan{\ss}en}, {Jevardat de
  Fombelle}, {Jordan}, {Krone-Martins}, {Lanzafame}, {L{\"o}ffler}, {Lorca},
  {Manteiga}, {Marchal}, {Marrese}, {Moitinho}, {Mora}, {Muinonen}, {Osborne},
  {Pancino}, {Pauwels}, {Petit}, {Recio-Blanco}, {Richards}, {Riello},
  {Rimoldini}, {Robin}, {Roegiers}, {Rybizki}, {Sarro}, {Siopis}, {Smith},
  {Sozzetti}, {Ulla}, {Utrilla}, {van Leeuwen}, {van Reeven}, {Abbas}, {Abreu
  Aramburu}, {Accart}, {Aerts}, {Aguado}, {Ajaj}, {Altavilla}, {{\'A}lvarez},
  {{\'A}lvarez Cid-Fuentes}, {Alves}, {Anderson}, {Anglada Varela}, {Antoja},
  {Audard}, {Baines}, {Baker}, {Balaguer-N{\'u}{\~n}ez}, {Balbinot}, {Balog},
  {Barache}, {Barbato}, {Barros}, {Barstow}, {Bartolom{\'e}}, {Bassilana},
  {Bauchet}, {Baudesson-Stella}, {Becciani}, {Bellazzini}, {Bernet}, {Bertone},
  {Bianchi}, {Blanco-Cuaresma}, {Boch}, {Bombrun}, {Bossini}, {Bouquillon},
  {Bragaglia}, {Bramante}, {Breedt}, {Bressan}, {Brouillet}, {Bucciarelli},
  {Burlacu}, {Busonero}, {Butkevich}, {Buzzi}, {Caffau}, {Cancelliere},
  {C{\'a}novas}, {Cantat-Gaudin}, {Carballo}, {Carlucci}, {Carnerero},
  {Carrasco}, {Casamiquela}, {Castellani}, {Castro-Ginard}, {Castro Sampol},
  {Chaoul}, {Charlot}, {Chemin}, {Chiavassa}, {Cioni}, {Comoretto}, {Cooper},
  {Cornez}, {Cowell}, {Crifo}, {Crosta}, {Crowley}, {Dafonte}, {Dapergolas},
  {David}, {David}, {de Laverny}, {De Luise}, {De March}, {De Ridder}, {de
  Souza}, {de Teodoro}, {de Torres}, {del Peloso}, {del Pozo}, {Delbo},
  {Delgado}, {Delgado}, {Delisle}, {Di Matteo}, {Diakite}, {Diener},
  {Distefano}, {Dolding}, {Eappachen}, {Edvardsson}, {Enke}, {Esquej}, {Fabre},
  {Fabrizio}, {Faigler}, {Fedorets}, {Fernique}, {Fienga}, {Figueras},
  {Fouron}, {Fragkoudi}, {Fraile}, {Franke}, {Gai}, {Garabato},
  {Garcia-Gutierrez}, {Garc{\'\i}a-Torres}, {Garofalo}, {Gavras}, {Gerlach},
  {Geyer}, {Giacobbe}, {Gilmore}, {Girona}, {Giuffrida}, {Gomel}, {Gomez},
  {Gonzalez-Santamaria}, {Gonz{\'a}lez-Vidal}, {Granvik},
  {Guti{\'e}rrez-S{\'a}nchez}, {Guy}, {Hauser}, {Haywood}, {Helmi}, {Hidalgo},
  {Hilger}, {H{\l}adczuk}, {Hobbs}, {Holland}, {Huckle}, {Jasniewicz},
  {Jonker}, {Juaristi Campillo}, {Julbe}, {Karbevska}, {Kervella}, {Khanna},
  {Kochoska}, {Kontizas}, {Kordopatis}, {Korn}, {Kostrzewa-Rutkowska},
  {Kruszy{\'n}ska}, {Lambert}, {Lanza}, {Lasne}, {Le Campion}, {Le Fustec},
  {Lebreton}, {Lebzelter}, {Leccia}, {Leclerc}, {Lecoeur-Taibi}, {Liao},
  {Licata}, {Lindstr{\o}m}, {Lister}, {Livanou}, {Lobel}, {Madrero Pardo},
  {Managau}, {Mann}, {Marchant}, {Marconi}, {Marcos Santos}, {Marinoni},
  {Marocco}, {Marshall}, {Martin Polo}, {Mart{\'\i}n-Fleitas}, {Masip},
  {Massari}, {Mastrobuono-Battisti}, {Mazeh}, {McMillan}, {Messina},
  {Michalik}, {Millar}, {Mints}, {Molina}, {Molinaro}, {Moln{\'a}r},
  {Montegriffo}, {Mor}, {Morbidelli}, {Morel}, {Morris}, {Mulone}, {Munoz},
  {Muraveva}, {Murphy}, {Musella}, {Noval}, {Ord{\'e}novic}, {Orr{\`u}},
  {Osinde}, {Pagani}, {Pagano}, {Palaversa}, {Palicio}, {Panahi}, {Pawlak},
  {Pe{\~n}alosa Esteller}, {Penttil{\"a}}, {Piersimoni}, {Pineau}, {Plachy},
  {Plum}, {Poggio}, {Poretti}, {Poujoulet}, {Pr{\v{s}}a}, {Pulone}, {Racero},
  {Ragaini}, {Rainer}, {Raiteri}, {Rambaux}, {Ramos}, {Ramos-Lerate}, {Re
  Fiorentin}, {Regibo}, {Reyl{\'e}}, {Ripepi}, {Riva}, {Rixon}, {Robichon},
  {Robin}, {Roelens}, {Rohrbasser}, {Romero-G{\'o}mez}, {Rowell}, {Royer},
  {Rybicki}, {Sadowski}, {Sagrist{\`a} Sell{\'e}s}, {Sahlmann}, {Salgado},
  {Salguero}, {Samaras}, {Sanchez Gimenez}, {Sanna}, {Santove{\~n}a},
  {Sarasso}, {Schultheis}, {Sciacca}, {Segol}, {Segovia}, {S{\'e}gransan},
  {Semeux}, {Shahaf}, {Siddiqui}, {Siebert}, {Siltala}, {Slezak}, {Smart},
  {Solano}, {Solitro}, {Souami}, {Souchay}, {Spagna}, {Spoto}, {Steele},
  {Steidelm{\"u}ller}, {Stephenson}, {S{\"u}veges}, {Szabados}, {Szegedi-Elek},
  {Taris}, {Tauran}, {Taylor}, {Teixeira}, {Thuillot}, {Tonello}, {Torra},
  {Torra}, {Turon}, {Unger}, {Vaillant}, {van Dillen}, {Vanel}, {Vecchiato},
  {Viala}, {Vicente}, {Voutsinas}, {Weiler}, {Wevers}, {Wyrzykowski}, {Yoldas},
  {Yvard}, {Zhao}, {Zorec}, {Zucker}, {Zurbach}, \& {Zwitter}}]{Gaia-EDR32020}
{Gaia Collaboration}, {Brown}, A.~G.~A., {Vallenari}, A., {et~al.} 2021,
  \href{http://dx.doi.org/10.1051/0004-6361/202039657}{\color{magenta}\aap},
  \href{https://ui.adsabs.harvard.edu/abs/2021A&A...649A...1G}{649, A1}

\bibitem[{{Galazutdinov} {et~al.}(2020){Galazutdinov}, {Bondar}, {Lee},
  {Hakalla}, {Szajna}, \& {Kre{\l}owski}}]{Galazutdinov2020}
{Galazutdinov}, G., {Bondar}, A., {Lee}, B.-C., {et~al.} 2020,
  \href{http://dx.doi.org/10.3847/1538-3881/ab6d01}{\color{magenta}\aj},
  \href{https://ui.adsabs.harvard.edu/abs/2020AJ....159..113G}{159, 113}

\bibitem[{{Galazutdinov} {et~al.}(2017{\natexlab{a}}){Galazutdinov}, {Lee},
  {Han}, {Lee}, {Valyavin}, \& {Kre{\l}owski}}]{Galazutdinov17a}
{Galazutdinov}, G.~A., {Lee}, J.-J., {Han}, I., {et~al.} 2017{\natexlab{a}},
  \href{http://dx.doi.org/10.1093/mnras/stx330}{\color{magenta}\mnras},
  \href{https://ui.adsabs.harvard.edu/abs/2017MNRAS.467.3099G}{467, 3099}

\bibitem[{{Galazutdinov} {et~al.}(2000){Galazutdinov}, {Musaev},
  {Kre{\l}owski}, \& {Walker}}]{Galazutdinov00}
{Galazutdinov}, G.~A., {Musaev}, F.~A., {Kre{\l}owski}, J., \& {Walker},
  G.~A.~H. 2000, \href{http://dx.doi.org/10.1086/316570}{\color{magenta}\pasp},
  \href{https://ui.adsabs.harvard.edu/abs/2000PASP..112..648G}{112, 648}

\bibitem[{{Galazutdinov} {et~al.}(2017{\natexlab{b}}){Galazutdinov},
  {Shimansky}, {Bondar}, {Valyavin}, \& {Kre{\l}owski}}]{Galazutdinov17b}
{Galazutdinov}, G.~A., {Shimansky}, V.~V., {Bondar}, A., {Valyavin}, G., \&
  {Kre{\l}owski}, J. 2017{\natexlab{b}},
  \href{http://dx.doi.org/10.1093/mnras/stw2948}{\color{magenta}\mnras},
  \href{https://ui.adsabs.harvard.edu/abs/2017MNRAS.465.3956G}{465, 3956}

\bibitem[{{Galazutdinov} {et~al.}(2021){Galazutdinov}, {Valyavin}, {Ikhsanov},
  \& {Kre{\l}owski}}]{Galazutdinov2021}
{Galazutdinov}, G.~A., {Valyavin}, G., {Ikhsanov}, N.~R., \& {Kre{\l}owski}, J.
  2021, \href{http://dx.doi.org/10.3847/1538-3881/abd4e5}{\color{magenta}\aj},
  \href{https://ui.adsabs.harvard.edu/abs/2021AJ....161..127G}{161, 127}

\bibitem[{{Geary}(1975)}]{Geary75}
{Geary}, J.~C. 1975,
  \href{https://ui.adsabs.harvard.edu/abs/1975PhDT........53G}{{The Use of a
  Self-Scanning Silicon Photodiode Array for Astronomical Spectroscopy.}}, PhD
  thesis, THE UNIVERSITY OF ARIZONA.

\bibitem[{{Geballe}(2016)}]{Geballe16}
{Geballe}, T.~R. 2016, in Journal of Physics Conference Series, Vol. 728,
  Journal of Physics Conference Series,
  \href{https://ui.adsabs.harvard.edu/abs/2016JPhCS.728f2005G}{062005}

\bibitem[{{Gilmore} {et~al.}(2012){Gilmore}, {Randich}, {Asplund}, {Binney},
  {Bonifacio}, {Drew}, {Feltzing}, {Ferguson}, {Jeffries}, {Micela},
  {Negueruela}, {Prusti}, {Rix}, {Vallenari}, {Alfaro}, {Allende-Prieto},
  {Babusiaux}, {Bensby}, {Blomme}, {Bragaglia}, {Flaccomio}, {Fran{\c{c}}ois},
  {Irwin}, {Koposov}, {Korn}, {Lanzafame}, {Pancino}, {Paunzen},
  {Recio-Blanco}, {Sacco}, {Smiljanic}, {Van Eck}, {Walton}, {Aden}, {Aerts},
  {Affer}, {Alcala}, {Altavilla}, {Alves}, {Antoja}, {Arenou}, {Argiroffi},
  {Asensio Ramos}, {Bailer-Jones}, {Balaguer-Nunez}, {Bayo}, {Barbuy},
  {Barisevicius}, {Barrado y Navascues}, {Battistini}, {Bellas Velidis},
  {Bellazzini}, {Belokurov}, {Bergemann}, {Bertelli}, {Biazzo}, {Bienayme},
  {Bland-Hawthorn}, {Boeche}, {Bonito}, {Boudreault}, {Bouvier}, {Brandao},
  {Brown}, {de Bruijne}, {Burleigh}, {Caballero}, {Caffau}, {Calura},
  {Capuzzo-Dolcetta}, {Caramazza}, {Carraro}, {Casagrande}, {Casewell},
  {Chapman}, {Chiappini}, {Chorniy}, {Christlieb}, {Cignoni}, {Cocozza},
  {Colless}, {Collet}, {Collins}, {Correnti}, {Covino}, {Crnojevic}, {Cropper},
  {Cunha}, {Damiani}, {David}, {Delgado}, {Duffau}, {Edvardsson}, {Eldridge},
  {Enke}, {Eriksson}, {Evans}, {Eyer}, {Famaey}, {Fellhauer}, {Ferreras},
  {Figueras}, {Fiorentino}, {Flynn}, {Folha}, {Franciosini}, {Frasca},
  {Freeman}, {Fremat}, {Friel}, {Gaensicke}, {Gameiro}, {Garzon}, {Geier},
  {Geisler}, {Gerhard}, {Gibson}, {Gomboc}, {Gomez}, {Gonzalez-Fernandez},
  {Gonzalez Hernandez}, {Gosset}, {Grebel}, {Greimel}, {Groenewegen},
  {Grundahl}, {Guarcello}, {Gustafsson}, {Hadrava}, {Hatzidimitriou}, {Hambly},
  {Hammersley}, {Hansen}, {Haywood}, {Heber}, {Heiter}, {Held}, {Helmi},
  {Hensler}, {Herrero}, {Hill}, {Hodgkin}, {Huelamo}, {Huxor}, {Ibata},
  {Jackson}, {de Jong}, {Jonker}, {Jordan}, {Jordi}, {Jorissen}, {Katz},
  {Kawata}, {Keller}, {Kharchenko}, {Klement}, {Klutsch}, {Knude}, {Koch},
  {Kochukhov}, {Kontizas}, {Koubsky}, {Lallement}, {de Laverny}, {van Leeuwen},
  {Lemasle}, {Lewis}, {Lind}, {Lindstrom}, {Lobel}, {Lopez Santiago}, {Lucas},
  {Ludwig}, {Lueftinger}, {Magrini}, {Maiz Apellaniz}, {Maldonado}, {Marconi},
  {Marino}, {Martayan}, {Martinez-Valpuesta}, {Matijevic}, {McMahon},
  {Messina}, {Meyer}, {Miglio}, {Mikolaitis}, {Minchev}, {Minniti}, {Moitinho},
  {Momany}, {Monaco}, {Montalto}, {Monteiro}, {Monier}, {Montes}, {Mora},
  {Moraux}, {Morel}, {Mowlavi}, {Mucciarelli}, {Munari}, {Napiwotzki},
  {Nardetto}, {Naylor}, {Naze}, {Nelemans}, {Okamoto}, {Ortolani}, {Pace},
  {Palla}, {Palous}, {Parker}, {Penarrubia}, {Pillitteri}, {Piotto}, {Posbic},
  {Prisinzano}, {Puzeras}, {Quirrenbach}, {Ragaini}, {Read}, {Read}, {Reyle},
  {De Ridder}, {Robichon}, {Robin}, {Roeser}, {Romano}, {Royer}, {Ruchti},
  {Ruzicka}, {Ryan}, {Ryde}, {Santos}, {Sanz Forcada}, {Sarro Baro},
  {Sbordone}, {Schilbach}, {Schmeja}, {Schnurr}, {Schoenrich}, {Scholz},
  {Seabroke}, {Sharma}, {De Silva}, {Smith}, {Solano}, {Sordo}, {Soubiran},
  {Sousa}, {Spagna}, {Steffen}, {Steinmetz}, {Stelzer}, {Stempels},
  {Tabernero}, {Tautvaisiene}, {Thevenin}, {Torra}, {Tosi}, {Tolstoy}, {Turon},
  {Walker}, {Wambsganss}, {Worley}, {Venn}, {Vink}, {Wyse}, {Zaggia},
  {Zeilinger}, {Zoccali}, {Zorec}, {Zucker}, {Zwitter}, \& {Gaia-ESO Survey
  Team}}]{Gilmore12}
{Gilmore}, G., {Randich}, S., {Asplund}, M., {et~al.} 2012, The Messenger,
  \href{https://ui.adsabs.harvard.edu/abs/2012Msngr.147...25G}{147, 25}

\bibitem[{{Gonzalez} {et~al.}(2011){Gonzalez}, {Rejkuba}, {Zoccali}, {Valenti},
  \& {Minniti}}]{Gonzalez11}
{Gonzalez}, O.~A., {Rejkuba}, M., {Zoccali}, M., {Valenti}, E., \& {Minniti},
  D. 2011,
  \href{http://dx.doi.org/10.1051/0004-6361/201117601}{\color{magenta}\aap},
  \href{https://ui.adsabs.harvard.edu/abs/2011A&A...534A...3G}{534, A3}

\bibitem[{{Gonzalez} {et~al.}(2012){Gonzalez}, {Rejkuba}, {Zoccali}, {Valenti},
  {Minniti}, {Schultheis}, {Tobar}, \& {Chen}}]{Gonzalez12}
{Gonzalez}, O.~A., {Rejkuba}, M., {Zoccali}, M., {et~al.} 2012,
  \href{http://dx.doi.org/10.1051/0004-6361/201219222}{\color{magenta}\aap},
  \href{https://ui.adsabs.harvard.edu/abs/2012A&A...543A..13G}{543, A13}

\bibitem[{{Gustafsson} {et~al.}(2008){Gustafsson}, {Edvardsson}, {Eriksson},
  {J{\o}rgensen}, {Nordlund}, \& {Plez}}]{Gustafsson08}
{Gustafsson}, B., {Edvardsson}, B., {Eriksson}, K., {et~al.} 2008,
  \href{http://dx.doi.org/10.1051/0004-6361:200809724}{\color{magenta}\aap},
  \href{https://ui.adsabs.harvard.edu/abs/2008A&A...486..951G}{486, 951}

\bibitem[{{Hamano} {et~al.}(2015){Hamano}, {Kobayashi}, {Kondo}, {Ikeda},
  {Nakanishi}, {Yasui}, {Mizumoto}, {Matsunaga}, {Fukue}, {Mito}, {Yamamoto},
  {Izumi}, {Nakaoka}, {Kawanishi}, {Kitano}, {Otsubo}, {Kinoshita},
  {Kobayashi}, \& {Kawakita}}]{Hamano2015}
{Hamano}, S., {Kobayashi}, N., {Kondo}, S., {et~al.} 2015,
  \href{http://dx.doi.org/10.1088/0004-637X/800/2/137}{\color{magenta}\apj},
  \href{https://ui.adsabs.harvard.edu/abs/2015ApJ...800..137H}{800, 137}

\bibitem[{{Heger}(1922)}]{Heger1922}
{Heger}, M.~L. 1922, Lick Observatory Bulletin,
  \href{https://ui.adsabs.harvard.edu/abs/1922LicOB..10..146H}{10, 146}

\bibitem[{{Heiter} {et~al.}(2021){Heiter}, {Lind}, {Bergemann}, {Asplund},
  {Mikolaitis}, {Barklem}, {Masseron}, {de Laverny}, {Magrini}, {Edvardsson},
  {J{\"o}nsson}, {Pickering}, {Ryde}, {Bayo Ar{\'a}n}, {Bensby}, {Casey},
  {Feltzing}, {Jofr{\'e}}, {Korn}, {Pancino}, {Damiani}, {Lanzafame}, {Lardo},
  {Monaco}, {Morbidelli}, {Smiljanic}, {Worley}, {Zaggia}, {Randich}, \&
  {Gilmore}}]{Heiter2021}
{Heiter}, U., {Lind}, K., {Bergemann}, M., {et~al.} 2021,
  \href{http://dx.doi.org/10.1051/0004-6361/201936291}{\color{magenta}\aap},
  \href{https://ui.adsabs.harvard.edu/abs/2021A&A...645A.106H}{645, A106}

\bibitem[{{Herbig}(1975)}]{Herbig75}
{Herbig}, G.~H. 1975,
  \href{http://dx.doi.org/10.1086/153400}{\color{magenta}\apj},
  \href{https://ui.adsabs.harvard.edu/abs/1975ApJ...196..129H}{196, 129}

\bibitem[{{Herbig} \& {Leka}(1991)}]{HL1991}
{Herbig}, G.~H. \& {Leka}, K.~D. 1991,
  \href{http://dx.doi.org/10.1086/170708}{\color{magenta}\apj},
  \href{https://ui.adsabs.harvard.edu/abs/1991ApJ...382..193H}{382, 193}

\bibitem[{{Hobbs} {et~al.}(2009){Hobbs}, {York}, {Thorburn}, {Snow}, {Bishof},
  {Friedman}, {McCall}, {Oka}, {Rachford}, {Sonnentrucker}, \&
  {Welty}}]{Hobbs09}
{Hobbs}, L.~M., {York}, D.~G., {Thorburn}, J.~A., {et~al.} 2009,
  \href{http://dx.doi.org/10.1088/0004-637X/705/1/32}{\color{magenta}\apj},
  \href{https://ui.adsabs.harvard.edu/abs/2009ApJ...705...32H}{705, 32}

\bibitem[{{Howard} {et~al.}(2009){Howard}, {Rich}, {Clarkson}, {Mallery},
  {Kormendy}, {De Propris}, {Robin}, {Fux}, {Reitzel}, {Zhao}, {Kuijken}, \&
  {Koch}}]{Howard09}
{Howard}, C.~D., {Rich}, R.~M., {Clarkson}, W., {et~al.} 2009,
  \href{http://dx.doi.org/10.1088/0004-637X/702/2/L153}{\color{magenta}\apjl},
  \href{https://ui.adsabs.harvard.edu/abs/2009ApJ...702L.153H}{702, L153}

\bibitem[{{Istiqomah} {et~al.}(2020){Istiqomah}, {Puspitarini}, \&
  {Arifyanto}}]{Istiqomah2020}
{Istiqomah}, A.~N., {Puspitarini}, L., \& {Arifyanto}, M.~I. 2020, in Journal
  of Physics Conference Series, Vol. 1523, Journal of Physics Conference
  Series, \href{https://ui.adsabs.harvard.edu/abs/2020JPhCS1523a2009I}{012009}

\bibitem[{{Jenniskens} \& {Desert}(1994)}]{Jenniskens94}
{Jenniskens}, P. \& {Desert}, F.~X. 1994, \aaps,
  \href{https://ui.adsabs.harvard.edu/abs/1994A&AS..106...39J}{106, 39}

\bibitem[{{Kos} {et~al.}(2013){Kos}, {Zwitter}, {Grebel}, {Bienayme}, {Binney},
  {Bland-Hawthorn}, {Freeman}, {Gibson}, {Gilmore}, {Kordopatis}, {Navarro},
  {Parker}, {Reid}, {Seabroke}, {Siebert}, {Siviero}, {Steinmetz}, {Watson}, \&
  {Wyse}}]{Kos13}
{Kos}, J., {Zwitter}, T., {Grebel}, E.~K., {et~al.} 2013,
  \href{http://dx.doi.org/10.1088/0004-637X/778/2/86}{\color{magenta}\apj},
  \href{https://ui.adsabs.harvard.edu/abs/2013ApJ...778...86K}{778, 86}

\bibitem[{{Kos} {et~al.}(2014){Kos}, {Zwitter}, {Wyse}, {Bienaym{\'e}},
  {Binney}, {Bland-Hawthorn}, {Freeman}, {Gibson}, {Gilmore}, {Grebel},
  {Helmi}, {Kordopatis}, {Munari}, {Navarro}, {Parker}, {Reid}, {Seabroke},
  {Sharma}, {Siebert}, {Siviero}, {Steinmetz}, {Watson}, \& {Williams}}]{Kos14}
{Kos}, J., {Zwitter}, T., {Wyse}, R., {et~al.} 2014,
  \href{http://dx.doi.org/10.1126/science.1253171}{\color{magenta}Science},
  \href{https://ui.adsabs.harvard.edu/abs/2014Sci...345..791K}{345, 791}

\bibitem[{{Kre{\l}owski}(2018)}]{Krelowski18}
{Kre{\l}owski}, J. 2018,
  \href{http://dx.doi.org/10.1088/1538-3873/aabd69}{\color{magenta}\pasp},
  \href{https://ui.adsabs.harvard.edu/abs/2018PASP..130g1001K}{130, 071001}

\bibitem[{{Kre{\l}owski} {et~al.}(2016){Kre{\l}owski}, {Galazutdinov}, {Mulas},
  {Bondar}, {Musaev}, {Shapovalova}, {Cecchi-Pestellini}, {Beletsky}, \&
  {Lee}}]{Krelowski16}
{Kre{\l}owski}, J., {Galazutdinov}, G.~A., {Mulas}, G., {et~al.} 2016, \actaa,
  \href{https://ui.adsabs.harvard.edu/abs/2016AcA....66..391K}{66, 391}

\bibitem[{{Kurucz}(2005)}]{Kurucz2005}
{Kurucz}, R.~L. 2005, Memorie della Societa Astronomica Italiana Supplementi,
  \href{https://ui.adsabs.harvard.edu/abs/2005MSAIS...8...14K}{8, 14}

\bibitem[{{Lallement} {et~al.}(2018){Lallement}, {Cox}, {Cami}, {Smoker},
  {Farhang}, {Elyajouri}, {Cordiner}, {Linnartz}, {Smith}, {Ehrenfreund}, \&
  {Foing}}]{Lallement18}
{Lallement}, R., {Cox}, N.~L.~J., {Cami}, J., {et~al.} 2018,
  \href{http://dx.doi.org/10.1051/0004-6361/201832647}{\color{magenta}\aap},
  \href{https://ui.adsabs.harvard.edu/abs/2018A&A...614A..28L}{614, A28}

\bibitem[{{Lan} {et~al.}(2015){Lan}, {M{\'e}nard}, \& {Zhu}}]{Lan15}
{Lan}, T.-W., {M{\'e}nard}, B., \& {Zhu}, G. 2015,
  \href{http://dx.doi.org/10.1093/mnras/stv1519}{\color{magenta}\mnras},
  \href{https://ui.adsabs.harvard.edu/abs/2015MNRAS.452.3629L}{452, 3629}

\bibitem[{{Linnartz} {et~al.}(2020){Linnartz}, {Cami}, {Cordiner}, {Cox},
  {Ehrenfreund}, {Foing}, {Gatchell}, \& {Scheier}}]{Linnartz2020}
{Linnartz}, H., {Cami}, J., {Cordiner}, M., {et~al.} 2020,
  \href{http://dx.doi.org/10.1016/j.jms.2019.111243}{\color{magenta}Journal of
  Molecular Spectroscopy},
  \href{https://ui.adsabs.harvard.edu/abs/2020JMoSp.36711243L}{367, 111243}

\bibitem[{{Maier} {et~al.}(2004){Maier}, {Walker}, \& {Bohlender}}]{Maier04}
{Maier}, J.~P., {Walker}, G. A.~H., \& {Bohlender}, D.~A. 2004,
  \href{http://dx.doi.org/10.1086/381027}{\color{magenta}\apj},
  \href{https://ui.adsabs.harvard.edu/abs/2004ApJ...602..286M}{602, 286}

\bibitem[{{Majewski} {et~al.}(2017){Majewski}, {Schiavon}, {Frinchaboy},
  {Allende Prieto}, {Barkhouser}, {Bizyaev}, {Blank}, {Brunner}, {Burton},
  {Carrera}, {Chojnowski}, {Cunha}, {Epstein}, {Fitzgerald}, {Garc{\'\i}a
  P{\'e}rez}, {Hearty}, {Henderson}, {Holtzman}, {Johnson}, {Lam}, {Lawler},
  {Maseman}, {M{\'e}sz{\'a}ros}, {Nelson}, {Nguyen}, {Nidever}, {Pinsonneault},
  {Shetrone}, {Smee}, {Smith}, {Stolberg}, {Skrutskie}, {Walker}, {Wilson},
  {Zasowski}, {Anders}, {Basu}, {Beland}, {Blanton}, {Bovy}, {Brownstein},
  {Carlberg}, {Chaplin}, {Chiappini}, {Eisenstein}, {Elsworth}, {Feuillet},
  {Fleming}, {Galbraith-Frew}, {Garc{\'\i}a}, {Garc{\'\i}a-Hern{\'a}ndez},
  {Gillespie}, {Girardi}, {Gunn}, {Hasselquist}, {Hayden}, {Hekker}, {Ivans},
  {Kinemuchi}, {Klaene}, {Mahadevan}, {Mathur}, {Mosser}, {Muna}, {Munn},
  {Nichol}, {O'Connell}, {Parejko}, {Robin}, {Rocha-Pinto}, {Schultheis},
  {Serenelli}, {Shane}, {Silva Aguirre}, {Sobeck}, {Thompson}, {Troup},
  {Weinberg}, \& {Zamora}}]{Majewski2017}
{Majewski}, S.~R., {Schiavon}, R.~P., {Frinchaboy}, P.~M., {et~al.} 2017,
  \href{http://dx.doi.org/10.3847/1538-3881/aa784d}{\color{magenta}\aj},
  \href{https://ui.adsabs.harvard.edu/abs/2017AJ....154...94M}{154, 94}

\bibitem[{{Majewski} {et~al.}(2011){Majewski}, {Zasowski}, \&
  {Nidever}}]{Majewski2011}
{Majewski}, S.~R., {Zasowski}, G., \& {Nidever}, D.~L. 2011,
  \href{http://dx.doi.org/10.1088/0004-637X/739/1/25}{\color{magenta}\apj},
  \href{https://ui.adsabs.harvard.edu/abs/2011ApJ...739...25M}{739, 25}

\bibitem[{{Marigo} {et~al.}(2017){Marigo}, {Girardi}, {Bressan}, {Rosenfield},
  {Aringer}, {Chen}, {Dussin}, {Nanni}, {Pastorelli}, {Rodrigues}, {Trabucchi},
  {Bladh}, {Dalcanton}, {Groenewegen}, {Montalb{\'a}n}, \& {Wood}}]{Marigo17}
{Marigo}, P., {Girardi}, L., {Bressan}, A., {et~al.} 2017,
  \href{http://dx.doi.org/10.3847/1538-4357/835/1/77}{\color{magenta}\apj},
  \href{https://ui.adsabs.harvard.edu/abs/2017ApJ...835...77M}{835, 77}

\bibitem[{{McCall} {et~al.}(2010){McCall}, {Drosback}, {Thorburn}, {York},
  {Friedman}, {Hobbs}, {Rachford}, {Snow}, {Sonnentrucker}, \&
  {Welty}}]{McCall10}
{McCall}, B.~J., {Drosback}, M.~M., {Thorburn}, J.~A., {et~al.} 2010,
  \href{http://dx.doi.org/10.1088/0004-637X/708/2/1628}{\color{magenta}\apj},
  \href{https://ui.adsabs.harvard.edu/abs/2010ApJ...708.1628M}{708, 1628}

\bibitem[{{Merrill}(1930)}]{Merrill30}
{Merrill}, P.~W. 1930,
  \href{http://dx.doi.org/10.1086/143266}{\color{magenta}\apj},
  \href{https://ui.adsabs.harvard.edu/abs/1930ApJ....72...98M}{72, 98}

\bibitem[{{Merrill} \& {Wilson}(1938)}]{Merrill38}
{Merrill}, P.~W. \& {Wilson}, O.~C. 1938,
  \href{http://dx.doi.org/10.1086/143897}{\color{magenta}\apj},
  \href{https://ui.adsabs.harvard.edu/abs/1938ApJ....87....9M}{87, 9}

\bibitem[{{Minniti} {et~al.}(2017){Minniti}, {Lucas}, \& {VVV
  Team}}]{Minniti17}
{Minniti}, D., {Lucas}, P., \& {VVV Team}. 2017,
  \href{https://ui.adsabs.harvard.edu/abs/2017yCat.2348....0M}{VizieR Online
  Data Catalog, II/348}

\bibitem[{{Minniti} {et~al.}(2010){Minniti}, {Lucas}, {Emerson}, {Saito},
  {Hempel}, {Pietrukowicz}, {Ahumada}, {Alonso}, {Alonso-Garcia}, {Arias},
  {Bandyopadhyay}, {Barb{\'a}}, {Barbuy}, {Bedin}, {Bica}, {Borissova},
  {Bronfman}, {Carraro}, {Catelan}, {Clari{\'a}}, {Cross}, {de Grijs},
  {D{\'e}k{\'a}ny}, {Drew}, {Fari{\~n}a}, {Feinstein}, {Fern{\'a}ndez
  Laj{\'u}s}, {Gamen}, {Geisler}, {Gieren}, {Goldman}, {Gonzalez}, {Gunthardt},
  {Gurovich}, {Hambly}, {Irwin}, {Ivanov}, {Jord{\'a}n}, {Kerins}, {Kinemuchi},
  {Kurtev}, {L{\'o}pez-Corredoira}, {Maccarone}, {Masetti}, {Merlo},
  {Messineo}, {Mirabel}, {Monaco}, {Morelli}, {Padilla}, {Palma}, {Parisi},
  {Pignata}, {Rejkuba}, {Roman-Lopes}, {Sale}, {Schreiber}, {Schr{\"o}der},
  {Smith}, {}, {Soto}, {Tamura}, {Tappert}, {Thompson}, {Toledo}, {Zoccali}, \&
  {Pietrzynski}}]{Minniti10}
{Minniti}, D., {Lucas}, P.~W., {Emerson}, J.~P., {et~al.} 2010,
  \href{http://dx.doi.org/10.1016/j.newast.2009.12.002}{\color{magenta}\na},
  \href{https://ui.adsabs.harvard.edu/abs/2010NewA...15..433M}{15, 433}

\bibitem[{{Monreal-Ibero} {et~al.}(2018){Monreal-Ibero}, {Weilbacher}, \&
  {Wendt}}]{Monreal-Ibero18}
{Monreal-Ibero}, A., {Weilbacher}, P.~M., \& {Wendt}, M. 2018,
  \href{http://dx.doi.org/10.1051/0004-6361/201732178}{\color{magenta}\aap},
  \href{https://ui.adsabs.harvard.edu/abs/2018A&A...615A..33M}{615, A33}

\bibitem[{{Monreal-Ibero} {et~al.}(2015){Monreal-Ibero}, {Weilbacher}, {Wendt},
  {Selman}, {Lallement}, {Brinchmann}, {Kamann}, \& {Sandin}}]{Monreal-Ibero15}
{Monreal-Ibero}, A., {Weilbacher}, P.~M., {Wendt}, M., {et~al.} 2015,
  \href{http://dx.doi.org/10.1051/0004-6361/201525854}{\color{magenta}\aap},
  \href{https://ui.adsabs.harvard.edu/abs/2015A&A...576L...3M}{576, L3}

\bibitem[{{Mr{\'o}z} {et~al.}(2019){Mr{\'o}z}, {Udalski}, {Skowron}, {Skowron},
  {Soszy{\'n}ski}, {Pietrukowicz}, {Szyma{\'n}ski}, {Poleski}, {Koz{\l}owski},
  \& {Ulaczyk}}]{Mroz19}
{Mr{\'o}z}, P., {Udalski}, A., {Skowron}, D.~M., {et~al.} 2019,
  \href{http://dx.doi.org/10.3847/2041-8213/aaf73f}{\color{magenta}\apjl},
  \href{https://ui.adsabs.harvard.edu/abs/2019ApJ...870L..10M}{870, L10}

\bibitem[{{Munari} {et~al.}(2008){Munari}, {Tomasella}, {Fiorucci},
  {Bienaym{\'e}}, {Binney}, {Bland-Hawthorn}, {Boeche}, {Campbell}, {Freeman},
  {Gibson}, {Gilmore}, {Grebel}, {Helmi}, {Navarro}, {Parker}, {Seabroke},
  {Siebert}, {Siviero}, {Steinmetz}, {Watson}, {Williams}, {Wyse}, \&
  {Zwitter}}]{Munari08}
{Munari}, U., {Tomasella}, L., {Fiorucci}, M., {et~al.} 2008,
  \href{http://dx.doi.org/10.1051/0004-6361:200810232}{\color{magenta}\aap},
  \href{https://ui.adsabs.harvard.edu/abs/2008A&A...488..969M}{488, 969}

\bibitem[{{Ness} {et~al.}(2013){Ness}, {Freeman}, {Athanassoula},
  {Wylie-de-Boer}, {Bland-Hawthorn}, {Asplund}, {Lewis}, {Yong}, {Lane},
  {Kiss}, \& {Ibata}}]{Ness13b}
{Ness}, M., {Freeman}, K., {Athanassoula}, E., {et~al.} 2013,
  \href{http://dx.doi.org/10.1093/mnras/stt533}{\color{magenta}\mnras},
  \href{https://ui.adsabs.harvard.edu/abs/2013MNRAS.432.2092N}{432, 2092}

\bibitem[{{Omont}(2016)}]{Omont2016}
{Omont}, A. 2016,
  \href{http://dx.doi.org/10.1051/0004-6361/201527685}{\color{magenta}\aap},
  \href{https://ui.adsabs.harvard.edu/abs/2016A&A...590A..52O}{590, A52}

\bibitem[{{Omont} {et~al.}(2019){Omont}, {Bettinger}, \&
  {T{\"o}nshoff}}]{Omont19}
{Omont}, A., {Bettinger}, H.~F., \& {T{\"o}nshoff}, C. 2019,
  \href{http://dx.doi.org/10.1051/0004-6361/201834953}{\color{magenta}\aap},
  \href{https://ui.adsabs.harvard.edu/abs/2019A&A...625A..41O}{625, A41}

\bibitem[{{Piecka} \& {Paunzen}(2020)}]{PP2020}
{Piecka}, M. \& {Paunzen}, E. 2020,
  \href{http://dx.doi.org/10.1093/mnras/staa1112}{\color{magenta}\mnras},
  \href{https://ui.adsabs.harvard.edu/abs/2020MNRAS.495.2035P}{495, 2035}

\bibitem[{{Plez}(2012)}]{Plez2012}
{Plez}, B. 2012, {Turbospectrum: Code for spectral synthesis}

\bibitem[{Puspitarini \& Lallement(2019)}]{PL2019}
Puspitarini, L. \& Lallement, R. 2019,
  \href{http://dx.doi.org/10.1088/1742-6596/1245/1/012027}{\color{magenta}Journal
  of Physics: Conference Series}, 1245, 1245

\bibitem[{{Puspitarini} {et~al.}(2015){Puspitarini}, {Lallement}, {Babusiaux},
  {Chen}, {Bonifacio}, {Sbordone}, {Caffau}, {Duffau}, {Hill}, {Monreal-Ibero},
  {Royer}, {Arenou}, {Peralta}, {Drew}, {Bonito}, {Lopez-Santiago}, {Alfaro},
  {Bensby}, {Bragaglia}, {Flaccomio}, {Lanzafame}, {Pancino}, {Recio-Blanco},
  {Smiljanic}, {Costado}, {Lardo}, {de Laverny}, \& {Zwitter}}]{Puspitarini15}
{Puspitarini}, L., {Lallement}, R., {Babusiaux}, C., {et~al.} 2015,
  \href{http://dx.doi.org/10.1051/0004-6361/201424391}{\color{magenta}\aap},
  \href{https://ui.adsabs.harvard.edu/abs/2015A&A...573A..35P}{573, A35}

\bibitem[{{Recio-Blanco} {et~al.}(2006){Recio-Blanco}, {Bijaoui}, \& {de
  Laverny}}]{Recio-Blanco2006}
{Recio-Blanco}, A., {Bijaoui}, A., \& {de Laverny}, P. 2006,
  \href{http://dx.doi.org/10.1111/j.1365-2966.2006.10455.x}{\color{magenta}\mnras},
  \href{https://ui.adsabs.harvard.edu/abs/2006MNRAS.370..141R}{370, 141}

\bibitem[{{Recio-Blanco} {et~al.}(2016){Recio-Blanco}, {de Laverny}, {Allende
  Prieto}, {Fustes}, {Manteiga}, {Arcay}, {Bijaoui}, {Dafonte}, {Ordenovic}, \&
  {Ordo{\~n}ez Blanco}}]{Recio-Blanco16}
{Recio-Blanco}, A., {de Laverny}, P., {Allende Prieto}, C., {et~al.} 2016,
  \href{http://dx.doi.org/10.1051/0004-6361/201425030}{\color{magenta}\aap},
  \href{https://ui.adsabs.harvard.edu/abs/2016A&A...585A..93R}{585, A93}

\bibitem[{{Reid} \& {Brunthaler}(2004)}]{RB04}
{Reid}, M.~J. \& {Brunthaler}, A. 2004,
  \href{http://dx.doi.org/10.1086/424960}{\color{magenta}\apj},
  \href{https://ui.adsabs.harvard.edu/abs/2004ApJ...616..872R}{616, 872}

\bibitem[{{Reid} {et~al.}(2019){Reid}, {Menten}, {Brunthaler}, {Zheng}, {Dame},
  {Xu}, {Li}, {Sakai}, {Wu}, {Immer}, {Zhang}, {Sanna}, {Moscadelli}, {Rygl},
  {Bartkiewicz}, {Hu}, {Quiroga-Nu{\~n}ez}, \& {van Langevelde}}]{Reid19}
{Reid}, M.~J., {Menten}, K.~M., {Brunthaler}, A., {et~al.} 2019,
  \href{http://dx.doi.org/10.3847/1538-4357/ab4a11}{\color{magenta}\apj},
  \href{https://ui.adsabs.harvard.edu/abs/2019ApJ...885..131R}{885, 131}

\bibitem[{{Reid} {et~al.}(2014){Reid}, {Menten}, {Brunthaler}, {Zheng}, {Dame},
  {Xu}, {Wu}, {Zhang}, {Sanna}, {Sato}, {Hachisuka}, {Choi}, {Immer},
  {Moscadelli}, {Rygl}, \& {Bartkiewicz}}]{Reid14}
{Reid}, M.~J., {Menten}, K.~M., {Brunthaler}, A., {et~al.} 2014,
  \href{http://dx.doi.org/10.1088/0004-637X/783/2/130}{\color{magenta}\apj},
  \href{https://ui.adsabs.harvard.edu/abs/2014ApJ...783..130R}{783, 130}

\bibitem[{{Rojas-Arriagada} {et~al.}(2017){Rojas-Arriagada}, {Recio-Blanco},
  {de Laverny}, {Mikolaitis}, {Matteucci}, {Spitoni}, {Schultheis}, {Hayden},
  {Hill}, {Zoccali}, {Minniti}, {Gonzalez}, {Gilmore}, {Randich}, {Feltzing},
  {Alfaro}, {Babusiaux}, {Bensby}, {Bragaglia}, {Flaccomio}, {Koposov},
  {Pancino}, {Bayo}, {Carraro}, {Casey}, {Costado}, {Damiani}, {Donati},
  {Franciosini}, {Hourihane}, {Jofr{\'e}}, {Lardo}, {Lewis}, {Lind}, {Magrini},
  {Morbidelli}, {Sacco}, {Worley}, \& {Zaggia}}]{Alvaro17}
{Rojas-Arriagada}, A., {Recio-Blanco}, A., {de Laverny}, P., {et~al.} 2017,
  \href{http://dx.doi.org/10.1051/0004-6361/201629160}{\color{magenta}\aap},
  \href{https://ui.adsabs.harvard.edu/abs/2017A&A...601A.140R}{601, A140}

\bibitem[{{Rojas-Arriagada} {et~al.}(2020){Rojas-Arriagada}, {Zasowski},
  {Schultheis}, {Zoccali}, {Hasselquist}, {Chiappini}, {Cohen}, {Cunha},
  {Fern{\'a}ndez-Trincado}, {Fragkoudi}, {Garc{\'\i}a-Hern{\'a}ndez},
  {Geisler}, {Gran}, {Lian}, {Majewski}, {Minniti}, {Monachesi}, {Nitschelm},
  \& {Queiroz}}]{Alvaro20}
{Rojas-Arriagada}, A., {Zasowski}, G., {Schultheis}, M., {et~al.} 2020,
  \href{http://dx.doi.org/10.1093/mnras/staa2807}{\color{magenta}\mnras},
  \href{https://ui.adsabs.harvard.edu/abs/2020MNRAS.499.1037R}{499, 1037}

\bibitem[{{Ruiz-Dern} {et~al.}(2018){Ruiz-Dern}, {Babusiaux}, {Arenou},
  {Turon}, \& {Lallement}}]{Ruiz-Dern18}
{Ruiz-Dern}, L., {Babusiaux}, C., {Arenou}, F., {Turon}, C., \& {Lallement}, R.
  2018,
  \href{http://dx.doi.org/10.1051/0004-6361/201731572}{\color{magenta}\aap},
  \href{https://ui.adsabs.harvard.edu/abs/2018A&A...609A.116R}{609, A116}

\bibitem[{{Sanner} {et~al.}(1978){Sanner}, {Snell}, \& {vanden
  Bout}}]{Sanner78}
{Sanner}, F., {Snell}, R., \& {vanden Bout}, P. 1978,
  \href{http://dx.doi.org/10.1086/156628}{\color{magenta}\apj},
  \href{https://ui.adsabs.harvard.edu/abs/1978ApJ...226..460S}{226, 460}

\bibitem[{{Sbordone}(2005)}]{Sbordone2005}
{Sbordone}, L. 2005, Memorie della Societa Astronomica Italiana Supplementi,
  \href{https://ui.adsabs.harvard.edu/abs/2005MSAIS...8...61S}{8, 61}

\bibitem[{{Sbordone} {et~al.}(2004){Sbordone}, {Bonifacio}, {Castelli}, \&
  {Kurucz}}]{Sbordone2004}
{Sbordone}, L., {Bonifacio}, P., {Castelli}, F., \& {Kurucz}, R.~L. 2004,
  Memorie della Societa Astronomica Italiana Supplementi,
  \href{https://ui.adsabs.harvard.edu/abs/2004MSAIS...5...93S}{5, 93}

\bibitem[{{Schlafly} \& {Finkbeiner}(2011)}]{SF11}
{Schlafly}, E.~F. \& {Finkbeiner}, D.~P. 2011,
  \href{http://dx.doi.org/10.1088/0004-637X/737/2/103}{\color{magenta}\apj},
  \href{https://ui.adsabs.harvard.edu/abs/2011ApJ...737..103S}{737, 103}

\bibitem[{{Schlegel} {et~al.}(1998){Schlegel}, {Finkbeiner}, \& {Davis}}]{SFD}
{Schlegel}, D.~J., {Finkbeiner}, D.~P., \& {Davis}, M. 1998,
  \href{http://dx.doi.org/10.1086/305772}{\color{magenta}\apj},
  \href{https://ui.adsabs.harvard.edu/abs/1998ApJ...500..525S}{500, 525}

\bibitem[{{Sch{\"o}nrich} {et~al.}(2010){Sch{\"o}nrich}, {Binney}, \&
  {Dehnen}}]{Schonrich10}
{Sch{\"o}nrich}, R., {Binney}, J., \& {Dehnen}, W. 2010,
  \href{http://dx.doi.org/10.1111/j.1365-2966.2010.16253.x}{\color{magenta}\mnras},
  \href{https://ui.adsabs.harvard.edu/abs/2010MNRAS.403.1829S}{403, 1829}

\bibitem[{{Shen} {et~al.}(2010){Shen}, {Rich}, {Kormendy}, {Howard}, {De
  Propris}, \& {Kunder}}]{Shen10}
{Shen}, J., {Rich}, R.~M., {Kormendy}, J., {et~al.} 2010,
  \href{http://dx.doi.org/10.1088/2041-8205/720/1/L72}{\color{magenta}\apjl},
  \href{https://ui.adsabs.harvard.edu/abs/2010ApJ...720L..72S}{720, L72}

\bibitem[{{Sonnentrucker} {et~al.}(2018){Sonnentrucker}, {York}, {Hobbs},
  {Welty}, {Friedman}, {Dahlstrom}, {Snow}, \& {York}}]{Sonnentrucker2018}
{Sonnentrucker}, P., {York}, B., {Hobbs}, L.~M., {et~al.} 2018,
  \href{http://dx.doi.org/10.3847/1538-4365/aad4a5}{\color{magenta}\apjs},
  \href{https://ui.adsabs.harvard.edu/abs/2018ApJS..237...40S}{237, 40}

\bibitem[{{Spitzer}(1978)}]{Spitzer78}
{Spitzer}, L. 1978, {Physical processes in the interstellar medium}

\bibitem[{{Steinmetz} {et~al.}(2006){Steinmetz}, {Zwitter}, {Siebert},
  {Watson}, {Freeman}, {Munari}, {Campbell}, {Williams}, {Seabroke}, {Wyse},
  {Parker}, {Bienaym{\'e}}, {Roeser}, {Gibson}, {Gilmore}, {Grebel}, {Helmi},
  {Navarro}, {Burton}, {Cass}, {Dawe}, {Fiegert}, {Hartley}, {Russell},
  {Saunders}, {Enke}, {Bailin}, {Binney}, {Bland -Hawthorn}, {Boeche},
  {Dehnen}, {Eisenstein}, {Evans}, {Fiorucci}, {Fulbright}, {Gerhard},
  {Jauregi}, {Kelz}, {Mijovi{\'c}}, {Minchev}, {Parmentier}, {Pe{\~n}arrubia},
  {Quillen}, {Read}, {Ruchti}, {Scholz}, {Siviero}, {Smith}, {Sordo}, {Veltz},
  {Vidrih}, {von Berlepsch}, {Boyle}, \& {Schilbach}}]{Steinmetz06}
{Steinmetz}, M., {Zwitter}, T., {Siebert}, A., {et~al.} 2006,
  \href{http://dx.doi.org/10.1086/506564}{\color{magenta}\aj},
  \href{https://ui.adsabs.harvard.edu/abs/2006AJ....132.1645S}{132, 1645}

\bibitem[{{Surot} {et~al.}(2020){Surot}, {Valenti}, {Gonzalez}, {Zoccali},
  {S{\"o}kmen}, {Hidalgo}, \& {Minniti}}]{Surot2020}
{Surot}, F., {Valenti}, E., {Gonzalez}, O.~A., {et~al.} 2020,
  \href{http://dx.doi.org/10.1051/0004-6361/202038346}{\color{magenta}\aap},
  \href{https://ui.adsabs.harvard.edu/abs/2020A&A...644A.140S}{644, A140}

\bibitem[{{Thorburn} {et~al.}(2003){Thorburn}, {Hobbs}, {McCall}, {Oka},
  {Welty}, {Friedman}, {Snow}, {Sonnentrucker}, \& {York}}]{Thorburn03}
{Thorburn}, J.~A., {Hobbs}, L.~M., {McCall}, B.~J., {et~al.} 2003,
  \href{http://dx.doi.org/10.1086/345665}{\color{magenta}\apj},
  \href{https://ui.adsabs.harvard.edu/abs/2003ApJ...584..339T}{584, 339}

\bibitem[{{Tielens}(2014)}]{Tielens14}
{Tielens}, A.~G.~G.~M. 2014, in The Diffuse Interstellar Bands, ed. J.~{Cami}
  \& N.~L.~J. {Cox}, Vol. 297,
  \href{https://ui.adsabs.harvard.edu/abs/2014IAUS..297..399T}{399--411}

\bibitem[{{Walker} {et~al.}(2017){Walker}, {Campbell}, {Maier}, \&
  {Bohlender}}]{Walker2017}
{Walker}, G.~A.~H., {Campbell}, E.~K., {Maier}, J.~P., \& {Bohlender}, D. 2017,
  \href{http://dx.doi.org/10.3847/1538-4357/aa77f9}{\color{magenta}\apj},
  \href{https://ui.adsabs.harvard.edu/abs/2017ApJ...843...56W}{843, 56}

\bibitem[{{Walker} {et~al.}(2016){Walker}, {Campbell}, {Maier}, {Bohlender}, \&
  {Malo}}]{Walker2016}
{Walker}, G.~A.~H., {Campbell}, E.~K., {Maier}, J.~P., {Bohlender}, D., \&
  {Malo}, L. 2016,
  \href{http://dx.doi.org/10.3847/0004-637X/831/2/130}{\color{magenta}\apj},
  \href{https://ui.adsabs.harvard.edu/abs/2016ApJ...831..130W}{831, 130}

\bibitem[{{Wallerstein} {et~al.}(2007){Wallerstein}, {Sandstrom}, \&
  {Gredel}}]{Wallerstein07}
{Wallerstein}, G., {Sandstrom}, K., \& {Gredel}, R. 2007,
  \href{http://dx.doi.org/10.1086/521835}{\color{magenta}\pasp},
  \href{https://ui.adsabs.harvard.edu/abs/2007PASP..119.1268W}{119, 1268}

\bibitem[{{Wang} {et~al.}(2021){Wang}, {Zhang}, {Huang}, {Chen}, {Wang}, \&
  {Wang}}]{Wang2021}
{Wang}, F., {Zhang}, H.~W., {Huang}, Y., {et~al.} 2021,
  \href{http://dx.doi.org/10.1093/mnras/stab848}{\color{magenta}\mnras},
  \href{https://ui.adsabs.harvard.edu/abs/2021MNRAS.504..199W}{504, 199}

\bibitem[{{Wenger} {et~al.}(2018){Wenger}, {Balser}, {Anderson}, \&
  {Bania}}]{Wenger18}
{Wenger}, T.~V., {Balser}, D.~S., {Anderson}, L.~D., \& {Bania}, T.~M. 2018,
  \href{http://dx.doi.org/10.3847/1538-4357/aaaec8}{\color{magenta}\apj},
  \href{https://ui.adsabs.harvard.edu/abs/2018ApJ...856...52W}{856, 52}

\bibitem[{{Xiang} {et~al.}(2017){Xiang}, {Li}, \& {Zhong}}]{XiangFY17}
{Xiang}, F.~Y., {Li}, A., \& {Zhong}, J.~X. 2017,
  \href{http://dx.doi.org/10.3847/1538-4357/835/1/107}{\color{magenta}\apj},
  \href{https://ui.adsabs.harvard.edu/abs/2017ApJ...835..107X}{835, 107}

\bibitem[{{Zack} \& {Maier}(2014)}]{Zack14}
{Zack}, L.~N. \& {Maier}, J.~P. 2014, in The Diffuse Interstellar Bands, ed.
  J.~{Cami} \& N.~L.~J. {Cox}, Vol. 297,
  \href{https://ui.adsabs.harvard.edu/abs/2014IAUS..297..237Z}{237--246}

\bibitem[{{Zasowski} {et~al.}(2015){Zasowski}, {M{\'e}nard}, {Bizyaev},
  {Garc{\'\i}a-Hern{\'a}ndez}, {Garc{\'\i}a P{\'e}rez}, {Hayden}, {Holtzman},
  {Johnson}, {Kinemuchi}, {Majewski}, {Nidever}, {Shetrone}, \&
  {Wilson}}]{Zasowski15}
{Zasowski}, G., {M{\'e}nard}, B., {Bizyaev}, D., {et~al.} 2015,
  \href{http://dx.doi.org/10.1088/0004-637X/798/1/35}{\color{magenta}\apj},
  \href{https://ui.adsabs.harvard.edu/abs/2015ApJ...798...35Z}{798, 35}

\bibitem[{{Zhao} {et~al.}(2021){Zhao}, {Schultheis}, {Recio-Blanco},
  {Kordopatis}, {de Laverny}, {Rojas-Arriagada}, {Zoccali}, {Surot}, \&
  {Valenti}}]{hz21}
{Zhao}, H., {Schultheis}, M., {Recio-Blanco}, A., {et~al.} 2021,
  \href{http://dx.doi.org/10.1051/0004-6361/202039736}{\color{magenta}\aap},
  \href{https://ui.adsabs.harvard.edu/abs/2021A&A...645A..14Z}{645, A14}

\bibitem[{{Zoccali} {et~al.}(2014){Zoccali}, {Gonzalez}, {Vasquez}, {Hill},
  {Rejkuba}, {Valenti}, {Renzini}, {Rojas-Arriagada}, {Martinez-Valpuesta},
  {Babusiaux}, {Brown}, {Minniti}, \& {McWilliam}}]{Zoccali14}
{Zoccali}, M., {Gonzalez}, O.~A., {Vasquez}, S., {et~al.} 2014,
  \href{http://dx.doi.org/10.1051/0004-6361/201323120}{\color{magenta}\aap},
  \href{https://ui.adsabs.harvard.edu/abs/2014A&A...562A..66Z}{562, A66}

\end{thebibliography}

\clearpage

\end{CJK*}

\end{document}